\documentclass{aa}  
\usepackage{txfonts}
\usepackage{hyperref}
\hypersetup{
    colorlinks=true,
    citecolor = blue,
    linkcolor=blue
            }
\usepackage{graphicx}
\usepackage{natbib}
\usepackage[dvipsnames]{xcolor}
\usepackage{booktabs}

\newcommand{\lama}{$\lambda$\,And}
\newcommand{\cahk}{Ca\,{\sc ii}\,H\&K}
\newcommand{\cairt}{Ca\,{\sc ii}\,IRT}
\newcommand{\Halpha}{H$\alpha$}

\newcommand{\kms}{km\,s$^{-1}$}
\newcommand{\ms}{m\,s$^{-1}$}
\newcommand{\ergs}{erg\,cm$^{-2}$s$^{-1}$}

\begin{document} 


   \title{Surface image and activity-corrected orbit of the RS\,CVn binary HR\,7275}
    \subtitle{Disentangling activity tracers}

\author{\"O. Adebali\inst{1,2}\thanks{Corresponding author; \texttt{oadebali@aip.de}},
M. Weber\inst{1}, K. G. Strassmeier\inst{1,2}, I. V. Ilyin\inst{1}, M. Steffen\inst{1}, \and Zs. K\H{o}v\'ari\inst{3,4}}

\institute{
Leibniz-Institute for Astrophysics Potsdam (AIP), An der Sternwarte 16, D-14482 Potsdam, Germany
\and
Institut f\"ur Physik und Astronomie, Universit\"at Potsdam, D-14476 Potsdam, Germany
\and
HUN-REN Research Centre for Astronomy and Earth Sciences, Konkoly Observatory, H-1121 Budapest, Hungary
\and
HUN-REN CSFK, MTA Centre of Excellence, H-1121 Budapest, Hungary
                }

   \date{Received: 16 October 2025 / Accepted: 10 December 2025}

 
  \abstract
  %
   {Quantifying stellar parameters and magnetic activity for cool stars in double-lined spectroscopic binaries (SB2) is not straightforward, as both stars contribute to the observed composite spectra and are likely variable. Disentangled component spectra allow a detailed analysis of a component's magnetic activity. 
    }
  %
   {We aim at separating the spectra of the two stellar components of the HR\,7275 SB2 system. Our further aim is a more accurate orbital solution by cleaning the observed radial velocities (RV) from activity perturbations of the spotted primary ("RV jitter") and obtain a surface image of this component.   
    }
  %
   {We provide time-series high- and ultra-high resolution optical spectra and apply two different disentangling methods. RV residuals are modeled with three-sine function fits. The primary's spectral-line profiles are modeled with the Doppler imaging code $i$MAP. Magnetic fields are measured for the primary based on least-square deconvolved Stokes-V line profiles. Chromospheric emission is determined from the line-cores of \cahk, \cairt\,8542\,\AA, and Balmer \Halpha. Before applying those analyses, we provide a disentangling technique to determine the system properties more accurately.
  }
  %
   {The Doppler image of the primary shows two large cool spots of size $\approx$20\% of the visible hemisphere plus three smaller spots, each still $\approx$13\% in size. In total, HR\,7275a exhibited an impressive spottedness of $\approx$40\%\ of its entire surface in May-June 2022. The RV is modulated by the rotation of the primary with maximum amplitudes of 320\,\ms\ and 650\,\ms\ for two different modulation behaviors during the 250\,d of our observations. This jitter is primarily caused by the varying asymmetries of the apparent disk brightness due to the cool spots. Its removal resulted in roughly ten times higher precision of the orbital elements. Our snapshot magnetic-field  measurements reveal phase-dependent (large-scale) surface fields between +0.6$\pm$2.0\,G at phase 0.1 and $-$15.2$\pm$2.7\,G at phase 0.6, indicating a complex magnetic morphology related to the location of the photospheric spots. We also obtain a logarithmic lithium abundance of 0.58$\pm$0.1 for HR\,7275a, indicating considerable mixing, and 0.16$^{+0.23}_{-0.63}$ for HR\,7275b, which is an extremely low value.  
    }
   {}


    \authorrunning{Adebali et al.}

\maketitle
\section{Introduction}

Starspots are important indicators of stellar activity, providing information about the underlying magnetic dynamo in late-type stars. These features allow us to track the operation of magnetic activity, as well as activity processes associated with different atmospheric layers (such as the photosphere and chromosphere). 

Active components with huge starspots in RS\,CVn type binary systems offer an excellent opportunity to monitor such activity processes. These systems were first described and studied in depth by \citet{Hall1972}, and then \citet{Vogt1983} applied the groundbreaking Doppler imaging technique to cool stars, opening up entirely new horizons. The first detailed catalog of active binary stars was compiled by \citet{Strassmeier1993}.
The study of these RS\,CVn stars has yielded several fundamental results regarding the relationship between photospheric starspots and global magnetic activity \citep[e.g.,][]{Strassmeier2009}. Continuing the series, in our paper we take a closer look at the activity of another RS\,CVn system, HR\,7275, and attempt to separate the activity signals from the precise orbital solution.

HR\,7275 is a binary system with a tidally locked primary with an orbital and rotational period of $\approx$28\,d, which has been confirmed by various authors using photometric methods \citep{Fried1982, Strassmeier1989} and RV measurements \citep{Young1944, Eker1989, Osten1998, Medeiros1999}. It is a bright-star-catalog target with visual magnitude of 5.89 \citep{Ducati2002} that makes it easy to observe. The system is an SB2 with a K2IV-III type (sub)giant primary accompanied by a warm G type secondary. The mass of the primary was determined by \cite{Eker1989} to be between 1-3\,$\rm M_{\odot}$ and the stellar radius was given by \cite{Osten1998} as 6.2\,$\rm R_{\odot}$ with a surface temperature of 4500\,K. The rotational line broadening ($v\sin i$) was measured to about 15\,\kms \citep[e.g.,][]{Osten1998, Medeiros1999}. Initially considering the system as a single-lined binary (SB1), orbital solutions were given by \cite{Young1944} and \cite{Eker1989}. An SB2 solution was first reported by \cite{Osten1998} and then revised shortly afterwards by \cite{Medeiros1999}. Based on the solutions by \citet{Eker1989}, the secondary star has a mass of 0.9-1.1\,$M_{\odot}$ and a surface temperature of 5500\,K is found by \citet{Osten1998}.

In the case of an SB2 system like HR\,7275, spectral disentangling is the first step before determining precise stellar parameters. There are several different technical approaches for this purpose \citep[e.g.][]{Hensberge2008, Weber2011, Sablowski2019}. \citet{Osten1998} used a grid of template stars to fit the components of the HR\,7275 system in the combined spectrum. A similar technique, but with a synthetic spectral grid, was used by \citet{Weber2011} for Capella ($\alpha$\,Aur), a system of two G giants. An iterative technique, called the direct spectral decomposition, was developed by \citet{Folsom2010} and successfully applied to the eclipsing binary AR\,Aur. Another iterative method, the experiment of \citet{Kriskovics2013}, in which a spectral disentangling technique was combined with Doppler imaging in order to recover the surface temperature maps for both components of the SB2 pre-main-sequence system V824\,Ara should be mentioned.

TiO lines are often used to give estimates of starspot coverage and temperatures \citep{Neff1995}.
Using this method, \citet{O'neal1996} studied five active and evolved stars, including HR\,7275, for which they determined a spot temperature of 3600\,K with a filling factor of 31\%. However, starspots, in addition to being tracers of activity-related magnetic phenomena, can also degrade the accuracy of spectroscopic RV measurements. Moreover, in a recent work by \citet{Strassmeier2024}, it was shown that spots not only interfere with RV measurements, but also in the case of not very distant objects, such as the spotted giant XX\,Tri, significant rotation-induced stellar photocenter variations appear, which impose limitations, e.g., for astrometric-based exoplanet searches. Thus, for precise orbital calculations, these effects must be taken into account, especially when detecting a third companion, such as an exoplanet or a third body in a binary system. \citet{Saar&Donahue1997} showed that starspots can produce RV perturbations of up to 200\,\ms, while \citet{Hatzes2002} demonstrated that these perturbations correlate with $v\sin i$ (up to 13\,\kms\ in their simulations) and the spot-filling factor. \citet{Desort2007} also showed how a single spot on a Sun-like star affects the line bisector variations. While their simulations allowed only a single spot, \citet{Boisse2011} used a more sophisticated two-spot model with which they also performed inclination tests. Most recently, \citet{Zhao2023} have come up with a code called SOAP-GPU that allows the user to efficiently model stellar activity at the spectral level, even for complex activity region configurations. Finally, based on ultra-high resolution spectroscopic data of the RS\,CVn system \lama, we demonstrated the effect of starspots on RV measurements based on real observations, which we then took into account when correcting the orbital elements of the binary \citep{Adebali2025}.

The structure of this paper is as follows. In Sect.\,\ref{Observation_Reduction} we present our observations and the data reduction process. In Sect.\,\ref{Data_Analysis} our spectral disentangling routines are described in detail, along with the RV calculations and the starspot correction required to make the orbital solution more accurate. The process of Doppler imaging is presented in Sect.\,\ref{DI_section}, while the investigation of chromospheric activity is covered in Sect.\,\ref{Chromo_Sect}, together with obtained magnetic field measurements and lithium abundance analyses. We summarize and conclude on our findings in Sect.\,\ref{Summary}.

\section{Observations and data reduction}\label{Observation_Reduction}

\subsection{SES@STELLA}

By using the 1.2-m STELLA (STELLar Actvity) robotic telescope \citep{stella2004} with the STELLA Echelle Spectrograph (SES), we obtained in total 148 high-resolution ($R$=55,000) and high signal-to-noise ratio (S/N) spectra. Data with STELLA were autonomously collected once per clear night during the observing seasons 2021 and 2022. The SES is a fixed-format echelle spectrograph equipped with an e2v 4k$\times$4k CCD featuring 15 $\mu$m pixels and a fiber-fed structure. The spectral resolution with a sampling of three pixels per resolution element is achieved  through a 2-slice image slicer and a 67 $\mu$m octagonal fiber that projects to 3.8 arc seconds on the sky. The wavelength coverage is from 3900 to 8800$\AA$. This corresponds to an effective resolution of 110 m$\AA$, or 5.5 \kms\,at 6000$\AA$. Typical observations used an integration time of 280\,s, yielding an average S/N of about 400 per pixel. Although the spectrograph is housed thermally, it is not pressure stabilized, resulting in a radial-velocity stability of at best 30 \ms\,\citep{vpnep}.

The SES data reduction pipeline is based on IRAF and described in detail in \citet{stella10}. The CCD images were corrected for bad pixels and cosmic-ray impacts. Bias levels were removed by subtracting the average overscan from each image followed by the subtraction of the mean of the (already overscan subtracted) master bias frame. The target spectra were flattened by dividing by a nightly master flat. After
removal of the scattered light, the one-dimensional spectra were extracted using the standard IRAF optimal extraction routine. The blaze function was then removed
from the target spectra, followed by a wavelength calibration using consecutively recorded Th-Ar spectra. Finally, the extracted spectral orders were continuum normalized by dividing with a flux-normalized synthetic spectrum.

\subsection{PEPSI@VATT and PEPSI-POL@LBT}

The 1.8-m Vatican Advanced Technology Telescope (VATT) was used in combination with the Large Binocular Telescope's (LBT) Potsdam Echelle Polarimetric and Spectroscopic Instrument \citep[PEPSI;][]{Strassmeier2015}. PEPSI is a stabilized and fiber-fed echelle spectrograph having two arms (red and blue) with three cross dispersers (CD) for each arm. For HR7275, we employed only three of the six CDs (CD\,III+CD\,V and CD\,III+CD\,VI). Their wavelength regions covered were 4800-5440$\AA$ (CD\,III) and 6278-9067$\AA$ (CD\,V\&VI). In its VATT mode the spectrograph provides a two-pixel resolution of $R$=250,000. The instrument resides in the basement of the LBT with a fiber connection of 450\,m from the VATT. The VATT had been operated manually each night between UT~May~10, 2022 and June~14, 2022, and thus covered one full rotation of the target star sampled by 28 spectra.

High-resolution Stokes IQUV spectra of HR\,7275 were obtained on three separate nights, July~7, October~15, and October~22, 2024, using the PEPSI polarimeter with the 11.8-m LBT (PEPSI-POL). Circular polarization measurements were carried out with a super-achromatic quarter-wave retarder paired with a Foster prism serving as the polarizing beam-splitter. Observations were taken at two retarder positions, offset by 90$\degr$, and processed using the differential method outlined by \citet{ilyin2012}, producing the Stokes I and V spectra discussed in Sect.~\ref{Chromo_Sect}. These polarimetric observations achieved a resolving power of 130,000 (2-pixel sampling), with a combined S/N of up to 3700:1 per pixel in the red wavelengths of the Stokes-I spectra, based on six sub-exposures per dataset.

VATT and LBT PEPSI data were reduced using the SDS4PEPSI package (“Spectroscopic Data Systems for PEPSI”), which is based on \citet{ilyin2000} and further detailed in \citet{Strassmeier2018, Strassmeier2015}. The reduction procedure comprises several image-processing steps: starting with bias subtraction and variance estimation of the raw frames, followed by super-master flat fielding to correct CCD spatial noise and the subtraction of scattered light. The pipeline then identifies the echelle orders and computes the wavelength solution from ThAr calibration images. Afterward, the optimal extraction of the image slicers is performed, along with the removal of cosmic spikes. The extracted spectra are normalized using the master flat-field spectrum to eliminate both CCD fringes and the blaze function. Finally, a global 2D continuum fit is applied, and all orders are rectified into a one-dimensional spectrum.

\begin{figure}
    \centering
    \includegraphics[width=\columnwidth]{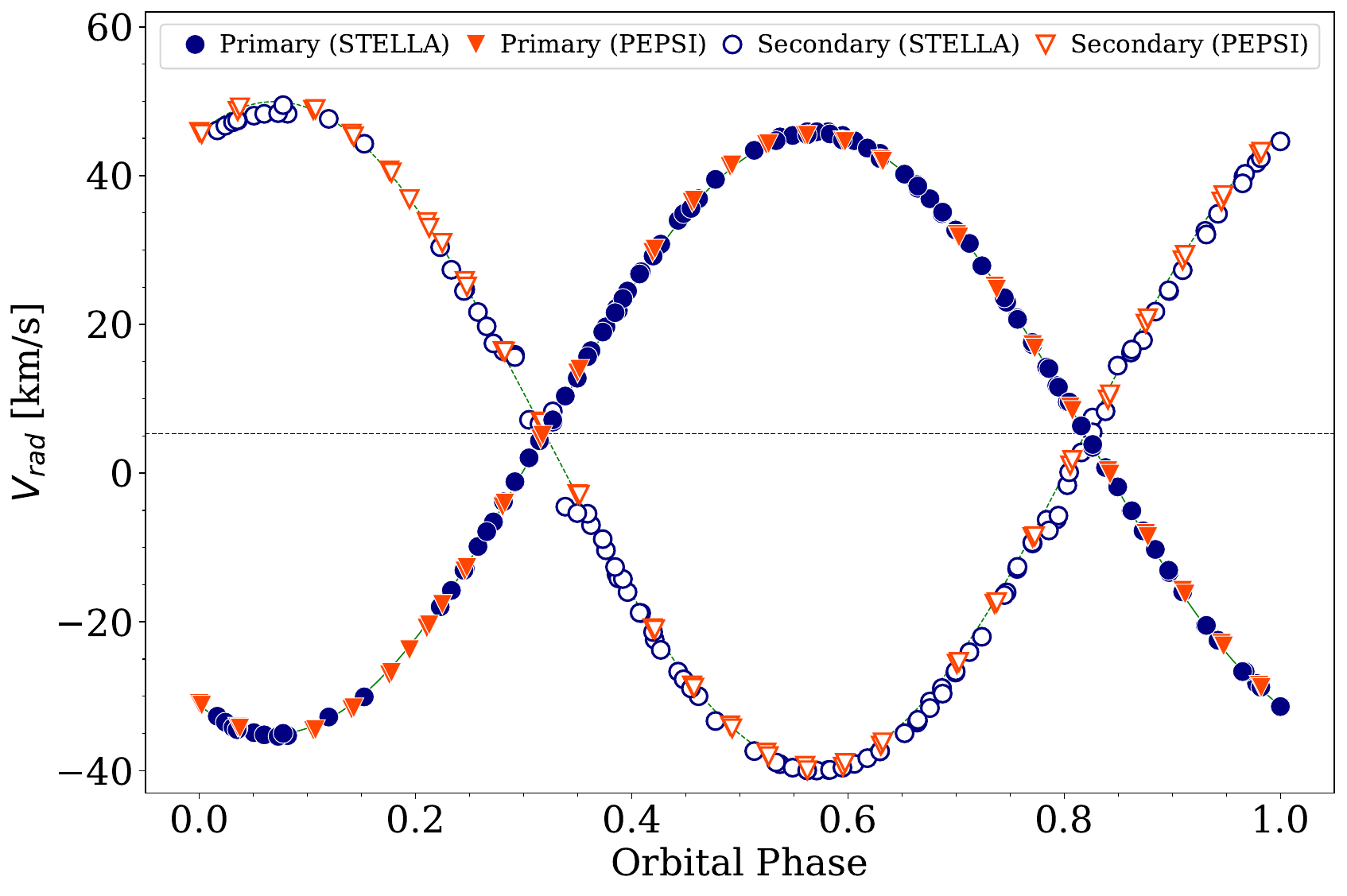}
    \caption{Phase folded radial velocities for both stars in the system. Circle data points indicate the STELLA observations; triangle data points show PEPSI observations. The green lines (solid and dashed) represent the orbital fit while the black dashed line is the center of mass velocity. }
    \label{RV_phase}
\end{figure}

\section{Data analysis}\label{Data_Analysis}

The observed spectra are analyzed using different techniques. First, we disentangle the two components to get a clearer picture of the activity behavior of the primary star. Then, we apply two different orbital solutions, both before and after the activity correction on the RV data points. The details of those processes are as follows.

\subsection{Spectral disentangling with \texttt{DISTRACT}}
 
Because HR\,7275 is an SB2 system, we search for the orbit of the secondary component by disentangling the spectra, using the code we developed for this purpose. \texttt{DISTRACT} (DIsentangler for STellaR ACTivity) consists of two different techniques for the two datasets with different spectral resolutions. For the lower spectral resolution STELLA data, we apply a two-dimensional cross-correlation technique to compute the SB2 solution. The higher-resolution PEPSI spectra are disentangled using a median subtraction technique. While ultimately we obtain a spot-corrected orbital solution with both approaches, the use of these two techniques allows us to obtain even more accurate results for the system's orbit and activity diagnostics.

\subsubsection{Two-dimensional cross-correlation}\label{2DCCF}

Considering the spectral resolution of the STELLA spectra, tracking the secondary star is possible using a model spectrum fitted to the observed spectra. For this purpose, we used two different MARCS template spectra with $T_{\rm eff}$=4500\,K, $\log g$=3.0 and [Fe/H]=$-$0.5 for the primary star and 5500\,K, 4.0, and 0.0 for the secondary, respectively \citep{Marcs2008}. We convolved the template spectra, while setting $v\sin i$ to 15\,\kms\ for the primary and 3\,\kms\ for the secondary. We then used a set of RV shifts between the templates to calculate the two-dimensional cross-correlation functions (2D-CCFs) for each velocity shift. The locations of the peak values of the one-dimensional projections of the 2D-CCF power spectrum indicate the different RV values of the two stars. These values are marked with blue circles along the orbital phase in Fig.\,\ref{RV_phase} (filled for the primary star, open for the secondary). More details about this technique can be found in \citet{Weber2011}.

\subsubsection{Median subtraction} \label{Med_Subt}

\begin{figure*}
    \centering
    \includegraphics[width=\textwidth]{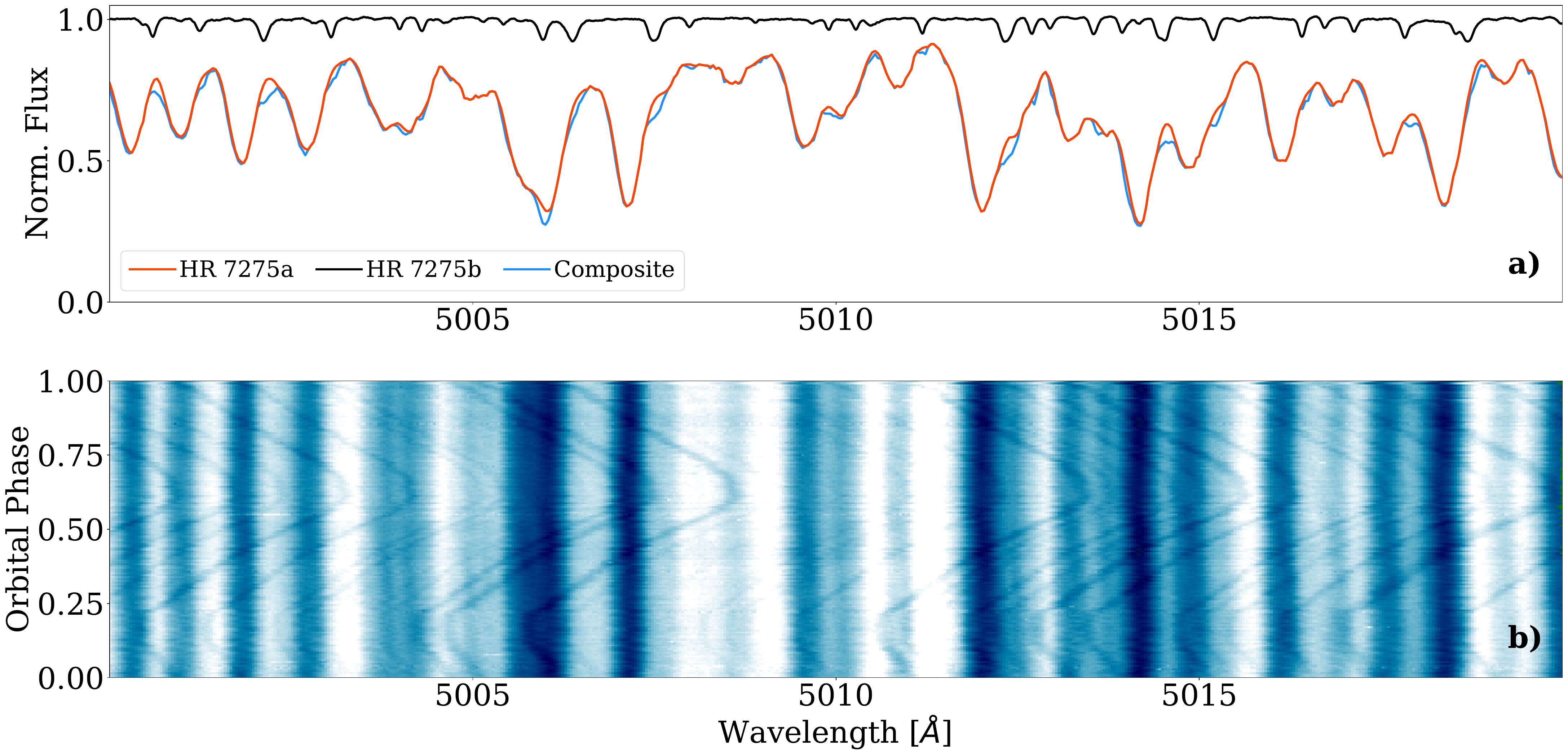}
    \caption{Spectral disentangling by median subtraction. Panel \emph{a}: The mean spectrum of the secondary star is plotted as a black line (top). The composite spectrum (blue) and a spectrum of the primary star (red) are plotted with a vertical shift of 0.15 at orbital phase 0.82. Panel \emph{b}: Time series composite spectra phase folded with the orbital period.}
    \label{Spec_phase}
\end{figure*}

The uniquely high resolution of the PEPSI spectra allows for easy detection of the secondary star. To this end, we arranged the observed spectra according to orbital phase and compiled the dynamic spectrum between 4800-5400\,$\AA$, which is shown in the lower panel of Fig.\,\ref{Spec_phase}. In the rest frame of the primary star, it is clearly visible how the lines of the secondary star shift according to the phase of the orbit. In order to separate the spectra of the two stars, we first prepared equally sampled time series composite spectra in the rest frame of the primary.

We then determined the median value of all observations for each wavelength value (essentially, therefore, the medians of the columns in the lower panel of Fig.\,\ref{Spec_phase}). This resulted in a single one-dimensional median composite spectrum. As a next step, we subtracted the median composite spectrum from each observed composite spectrum. This allowed us to obtain the spectrum of the secondary star for each observation time, taking into account the intrinsic variability of the primary star's activity. Considering the orbital velocity corrections, as a final step we averaged these median subtracted spectra, thus obtaining the averaged ("mean") spectrum of the secondary star. This spectrum is shown in Fig.\,\ref{hr7275b_spec} with a template spectrum of the secondary star. This technique for disentangling the stellar spectra has the advantage of being model-independent, or that it preserves the line depth ratio of the stars.

\begin{figure*}[h!!]
    \centering
    \includegraphics[width=\textwidth]{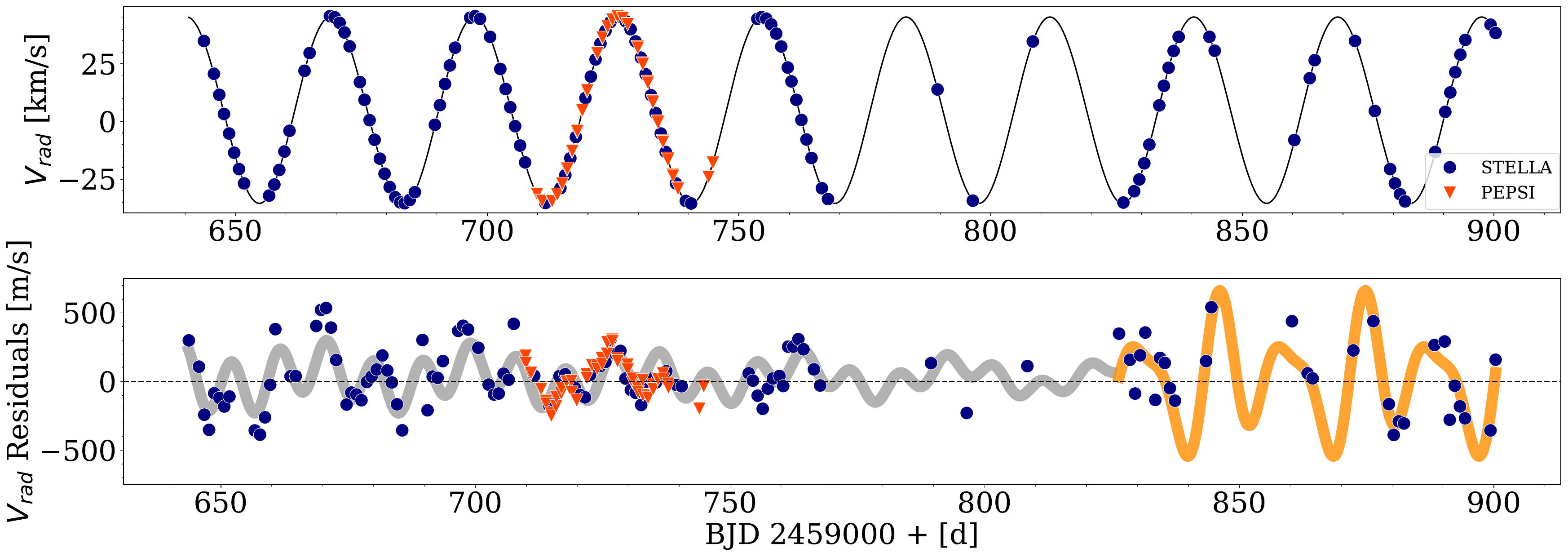}
    \caption{Radial velocities of HR\,7275a from VATT+PEPSI and STELLA+SES. In the upper panel, panel $a$, the SB1 orbital fit is plotted as a black line along with the data. STELLA+SES observations are indicated with dark blue circles; VATT+PEPSI observations are plotted in red triangles. In panel $b$ the RV residuals are shown after removing the predicted orbital velocities. The RV jitter appears multi-peaked per rotation. The thick grey and orange lines emphasize two different rotational modulation models.}
    \label{RV_fit}
\end{figure*}

\begin{figure}[h!]
    \centering
    \includegraphics[width=\columnwidth]{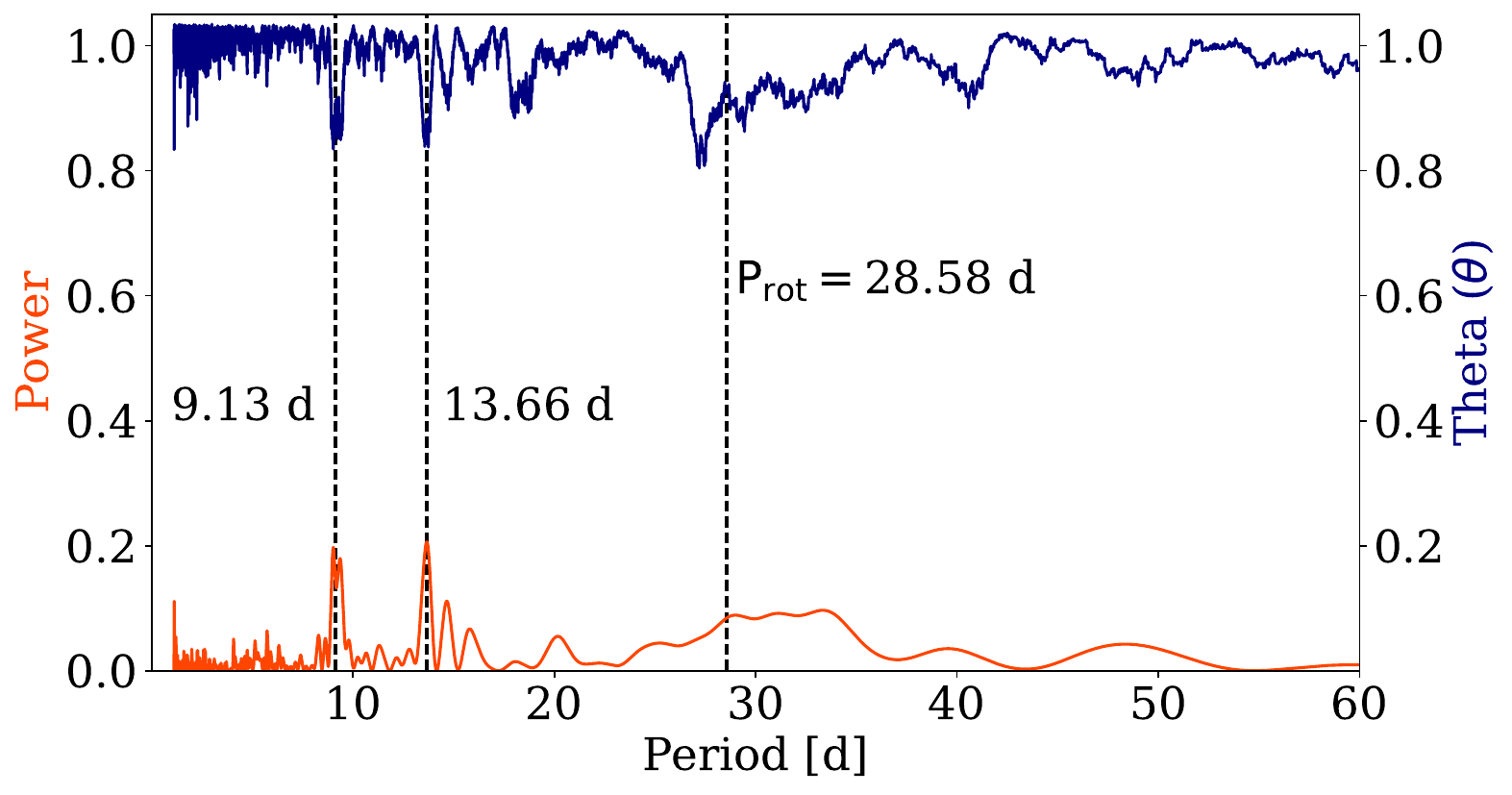}
    \caption{Lomb-Scargle periodogram (red) of RV residuals from Fig. \ref{RV_fit}. The strongest peak appears at the half rotation of the star. Also one third of the rotation appears prominently as the second strongest peak in the periodogram. Phase dispersion minimization (PDM) is shown in blue which shows the strongest peak at 27.2 days.}
    \label{LSP}
\end{figure}

\begin{table*}[ht!]
\caption{Orbital elements for HR\,7275}\label{hr7275_orb}
\begin{flushleft}
\begin{tabular}{lllll}
\hline\hline
\noalign{\smallskip}
Parameter           & SB2 solution,               & SB1 solution,               & \citet{Medeiros1999}       &\citet{Osten1998}\\
                    & spot-corrected         & spot-corrected            &                           & \\
\noalign{\smallskip}\hline\noalign{\smallskip}
$P_{\rm orb}$ [d]   & \multicolumn{2}{c}{28.586019 $\pm$ 0.000015}            & 28.5903 $\pm$ 0.0004        & 28.58973 $\pm$ 0.00002 \\
$K_{1}$ [\kms]          & 40.485 $\pm$ 0.006           & 40.428 $\pm$ 0.003         & 40.74 $\pm$ 0.16             & 40.15 $\pm$ 0.01 \\
$K_{2}$ [\kms]          & 44.715 $\pm$ 0.029           & \dots         & 45.05 $\pm$ 0.69     & 45.82 $\pm$ 0.01 \\
$\gamma$ [\kms]     & 4.902 $\pm$ 0.012           & 5.202 $\pm$ 0.005           & 4.11 $\pm$ 0.14             & 5.97 $\pm$ 0.19  \\
$T_0$ (BJD)         & 2459773.74 $\pm$ 0.12     & 2461403.02 $\pm$ 0.08   & 2448976.78 $\pm$ 0.05     & 2431043.908 $\pm$ 0.01  \\
$e$                 & 0.005 $\pm$ 0.001         & 0.0105 $\pm$ 0.0007        & 0.010 $\pm$ 0.004           & 0  \\
$\omega_{1}$ [deg]     & 239.46 $\pm$ 3.25             & 368.57 $\pm$ 1.52             & 0.00 $\pm$ 31.27           & \dots \\
$\omega_{2}$ [deg]     & 59.46 $\pm$ 3.25             & \dots              & \dots           & \dots \\
$a_1\sin i$ [Mkm] & 15.933 $\pm$ 0.023      & 15.890 $\pm$ 0.011         & 16.016 $\pm$ 0.064                     & 15.784 $\pm$ 0.004 \\
$a_2\sin i$ [Mkm] & 17.545 $\pm$ 0.021      & \dots         & \dots                     & 18.014 $\pm$ 0.004 \\
$\rm M_{1}$$sin^{3}i$ [$\rm M_{\odot}$]            & 0.9585 $\pm$ 0.0029     & \dots     & 0.9843 $\pm$ 0.0247                   & 1.0070 $\pm$ 0.0001 \\
$\rm M_{2}$$sin^{3}i$ [$\rm M_{\odot}$]            & 0.8702 $\pm$ 0.0028     & \dots     & 0.8901 $\pm$ 0.0186                   & 0.8823 $\pm$ 0.0001 \\
No. of obs.         & 176                      & 176                     & 28                        & 28 \\
Error of obs. of    & & & & \\
\ \ unit weight [\kms] & 0.490                   & 0.137                     & 0.673                      & \dots \\
\noalign{\smallskip}
\hline
\end{tabular}
\end{flushleft}
\end{table*}

\subsection{Further RV analysis}

During the disentangling process, we obtained the RV values of both stars. However, in order to track the activity changes of the primary star on the RV curve, we performed the following steps. From all observed spectra, we subtracted the mean secondary spectrum, shifted by the corresponding RV values for each observation time. Thus, we obtained the spectrum of the primary star for each observing day. The RV values of the primary star HR\,7275a were calculated by cross-correlating these spectra with the MARCS template spectrum assuming $T_{\rm eff}$=4500\,K, $\log g$=3.0, [Fe/H]=$-$0.5 and $v\sin i$ broadening of 16\,\kms. 

At this stage we must take into account the zero-point offset between the STELLA-SES and PEPSI instruments.
\citet{vpnep} found the SES-minus-PEPSI grand mean RV difference to be $-$395$\pm$209\,\ms. \citet{Adebali2025} found this difference to be $-$180\,\ms for the RS\,CVn system \lama. In this paper, we obtained the same difference of $-$180\,\ms\ for HR\,7275 by taking the difference of the mean values for this observing window, with which we corrected the SES RV values to match PEPSI. We note that the PEPSI data for \lama\ and HR\,7275 were obtained during the same nights close in time and are expected to be practically equal. Its uncertainty is estimated from the minimization of the rms of the orbital SB1 fit to be at most 137\,\ms. 

Since STELLA is a robotic telescope, observations were made every night according to the scheduled time, but some of the spectra had low S/N due to unfavorable weather conditions. Therefore, to eliminate outliers in the RV data set, a $\rm 3\sigma$ clipping was applied (14 low-S/N spectra were removed) . The resulting high-quality RV data were used in the following.

\subsection{Spot correction}

One of the major challenges for stars with large cool spots is determining the RV jitter caused by these dark features, which distort the observed spectra. Depending on the size of the starspot and its location on the stellar disk, the spot blocks the incident light beam and shifts the core of the spectral line. If the spot is located on the approaching half of the stellar disk, it blocks the light in such a way that a net redshift is observed, but if it is located on the receding half, the lack of light causes a net blueshift  \citep[see][]{Menuier2023}. These modulations produce a sinusoidal shape, which was modeled by \citet{Saar&Donahue1997} to quantify their effect on the RV curve. More complex models and observations have also been performed to obtain very accurate RV observations for the detection of exoplanets or long-term activity modulations on the surface of stars \citep[e.g.,][]{Boisse2011, Zhao2023, Adebali2025}.

HR\,7275a is very active not only compared to the Sun, but also compared to similar types of active stars, such as \lama\ that we recently studied \citep{Adebali2025}. Because of the large starspots, it is easier to track the spot-related modulations in its RV curve than for stars with smaller spots. Therefore, to track the effect of spots, we obtained an SB1 orbital solution for the primary from approximately nine orbital revolutions of RV data as shown in Fig.\,\ref{RV_fit}. The resulting observed-minus-calculated ($O-C$) residuals are shown in the bottom panel of the figure. The RV modulation depicted on the $O-C$ curve shows a complex modulation during the time scale of $\approx$250\,d. We see two trends with a separation by an observing gap of about 60 days (two stellar rotations): an epoch of lower jitter followed by an epoch of twice as large jitter. To model these two different amplitudes, we fitted three-sine functions with a small period shift which is typically caused by residual spot migration in latitude due to  differential surface rotation. This is introduced as a time-dependent period shift with a calculated maximum of $\sim$2 days during our observation window. The first global behavior is indicated in the bottom panel of Fig.\,\ref{RV_fit} with a gray thick line between BJD\,2,459,640-828. During this observing window, we calculated a RV modulation of up to 320\,\ms\ with the stellar rotation period of 28.58$\pm$1.3\,d was seen \citep[e.g.,][]{Fried1982, Strassmeier1989}. The periodicity clearly indicates a global cause of the activity-induced RV variability. The second behavior between BJD\,2,459,828-900 is indicated in the bottom panel of Fig.\,\ref{RV_fit} with a thick orange line and modeled in the same way. Although the data points are distributed more sparsely, we still detect the same period of $28.62\pm 1.9$\,d but with a much higher RV amplitude for the second global behavior with value of 650\,\ms. Both calculations are done with the minimum $\chi^{2}$ approach of \texttt{scipy}\ package \citep{SciPy2020-NMeth}. To remove these spot effects from the final RV model, we subtract these fitted residual functions from the observed RV data.

Figure\,\ref{LSP} shows the Lomb-Scargle periodogram of the fitted RV residuals, indicating the strongest peak at exactly half the rotation period. Another peak is found at one-third of the stellar rotation. However, this phenomenon is not unusual for stars with multiple spot configurations \citep[see e.g.,][]{Boisse2011, Adebali2025} and can be seen not only in RV residuals but also in photometric modulations \citep[e.g.,][]{Kovari2013}.

\subsection{Spot-corrected orbital solutions}

We use both SB1 and SB2 configurations for our orbital solutions. For the eccentric anomaly in both configurations, we follow the recipe published by \citet{dan:bur}. For the Keplerian orbit fitting, we used the \texttt{Python-scipy} implementation of a least-square algorithm. The procedure is the following: We first determined an SB1 orbit using all available data from \citet{Young1944}, \citet{Eker1989} and \citet{Osten1998}, spanning over 80 years. From this calculation, we kept only the orbital period and then fixed this value. This was followed by determining an  SB2 solution. We then also calculated a spot-corrected SB1 orbit for the primary. The final spot-corrected solutions are listed in Table\,\ref{hr7275_orb}, along with the best solutions taken from the literature. HR\,7275 exhibits a stable, synchronized, and nearly circular orbital configuration. We note that although the secondary star may also have spots, their effect is essentially negligible compared to that of the K2IV-III primary due to the difference in brightness and the star's low $v\sin i$ of $\approx$1-2\,\kms. We thus did not calculate a spot correction for the secondary. 

Our initial spot-uncorrected SB2 solution already gave significantly better fits than the most recent orbital calculation by \citet{Medeiros1999}. After spot correction, the error for an observation of unit weight decreased further from 0.51\,\kms\ to 0.49\,\kms for the SB2 solution, and from 0.22 to 0.14 for the SB1 solution. The errors of the orbital elements on average decreased by almost a factor ten compared to \citet{Medeiros1999} which favored a marginal non-circular orbit with a low eccentricity of 0.005$\pm$0.001. Although the surface spots of HR\,7275a change slowly with each rotation, the spot correction still allows an orbital solution for the primary alone. It is more than twice as accurate as the SB2 solution due to the more precise RV measurements for the primary with respect to the faint secondary and is also listed in Table\,\ref{hr7275_orb}.

\section{Doppler imaging} \label{DI_section}

\subsection{Data input and code summary}

The surface map of HR\,7275a is reconstructed using the $i$MAP code \citep{carroll12}, employing a multi-line inversion approach based on an average spectral line constructed from more than 500 individual lines with line depths exceeding 60\,\%\ of the continuum. The rotational line broadening of 15.4\,\kms of HR\,7275a makes it a difficult but possible Doppler imaging (DI) target. Atomic data for the multi-line inversion are drawn from the Vienna Atomic Line Database VALD-3 \citep{vald3}, covering the 4800–5400\,\AA\ wavelength range of our PEPSI CD\,III spectra. We preferred the 4800-5400\,\AA\ range for DI because we have CD\,III exposures for every observing night and thus the best phase coverage when compared to the red CDs. The (pseudo) average spectral line is computed using a Singular Value Decomposition (SVD) algorithm, where the dimensionality (rank) of the signal subspace was determined via a bootstrap-permutation test \citep[see][for details]{carroll12}. Noise estimates are similarly obtained from the bootstrap procedure. This methodology results in weighted mean line profiles with typical S/N of approximately 20,000 per pixel, substantially higher than the $\simeq$\,400–900 S/N per pixel achieved for the individual spectra.

\begin{figure*}[h!]
    \centering
    \includegraphics[width=\textwidth]{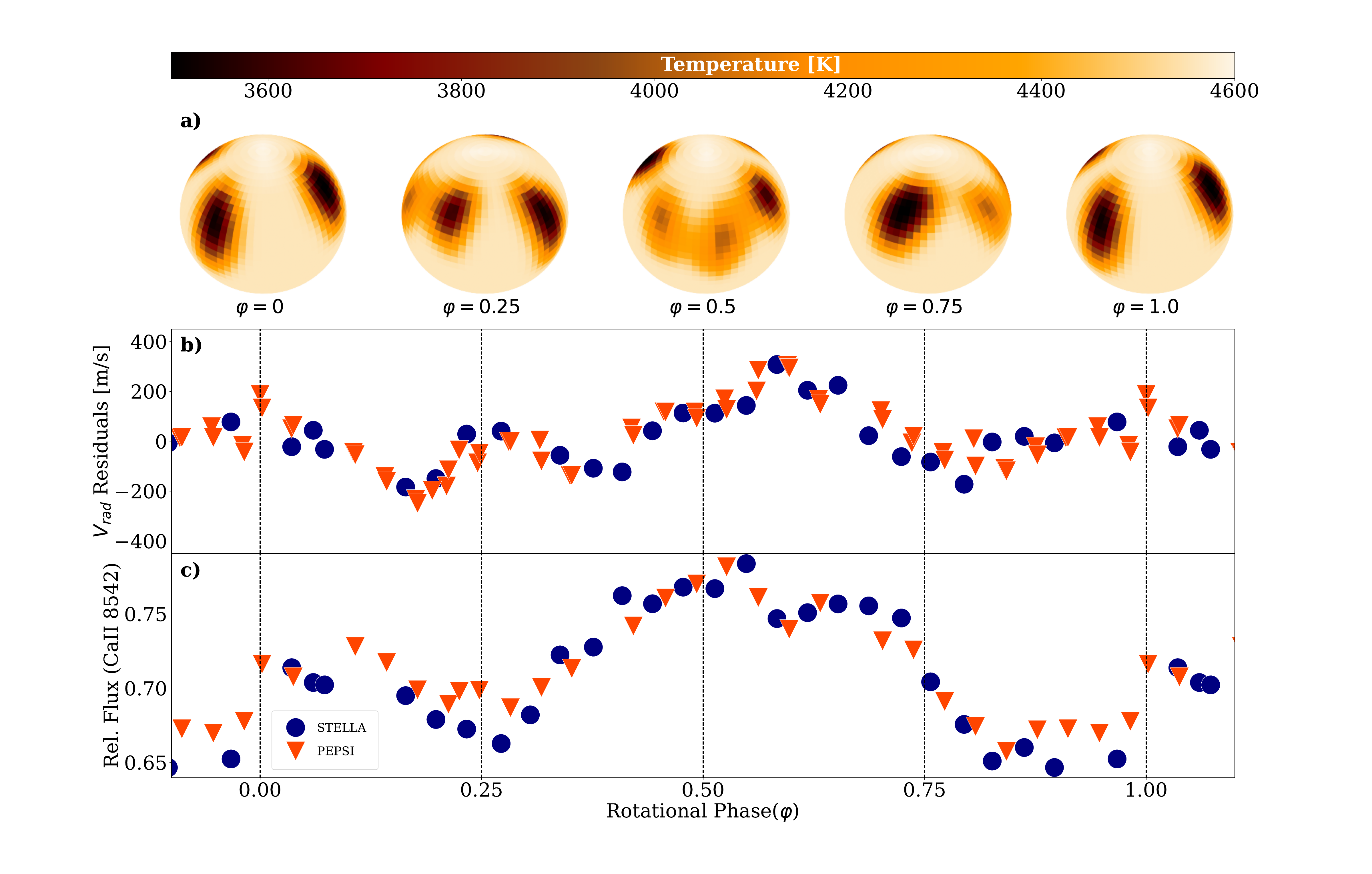}
    \caption{Panel \emph{a}: Doppler image of HR\,7275a. The rotational phases are indicated via phase $\varphi$ with a sampling of 0.25. Panel \emph{b}: Contemporaneous RV modulation due to the spot modulation. Panel \emph{c}: Relative flux modulation of \cairt\,8542\,$\AA$ simultaneous to the DI. Dark blue dots show the STELLA data set and red triangles indicate the PEPSI observations.}
    \label{DI_HR7275}
\end{figure*}

\begin{figure}[h!]
    \centering
    \includegraphics[width=\columnwidth]{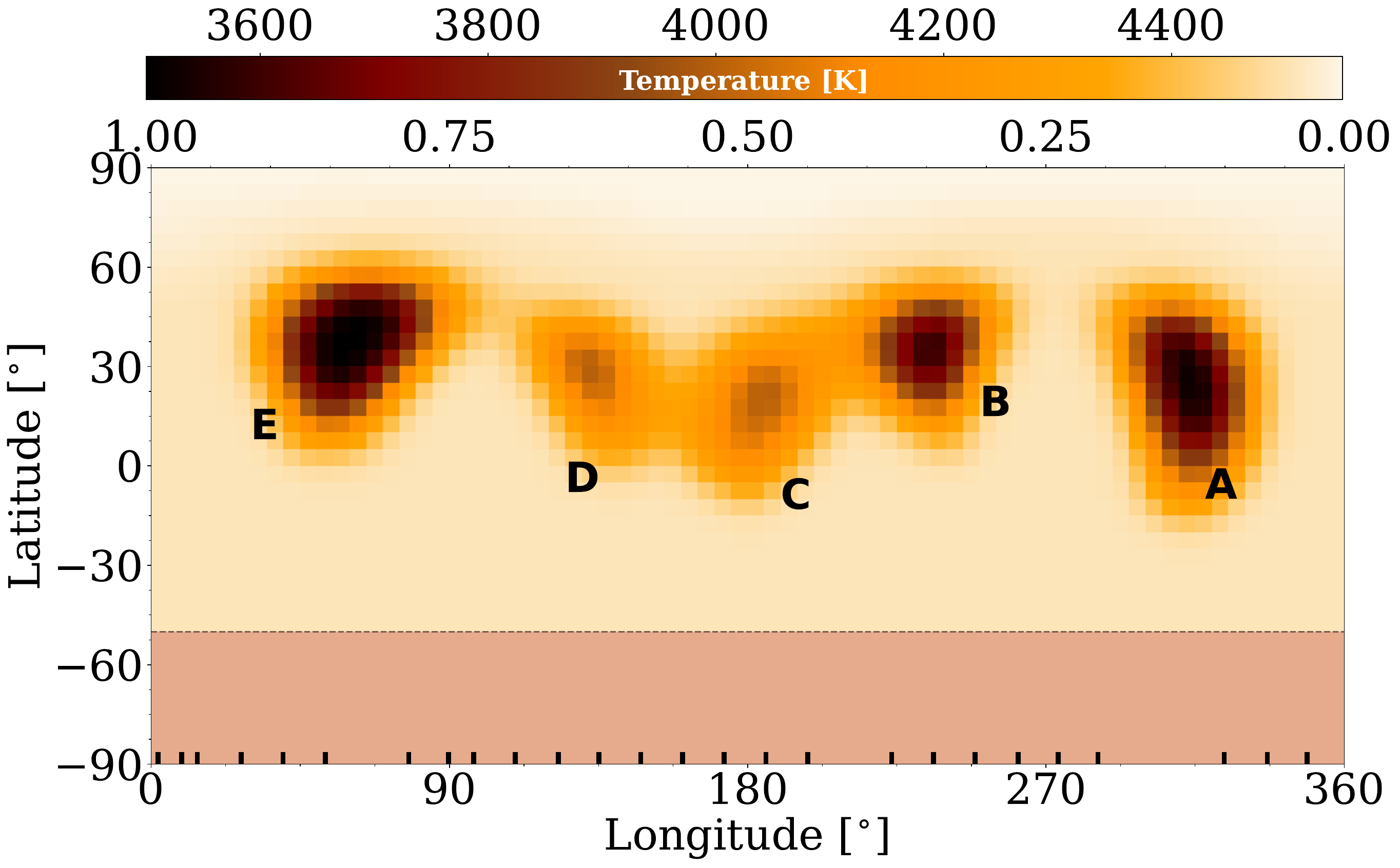}
    \caption{Mercator map of the spotted surface of HR 7275a. }
    \label{Mercator_DI}
\end{figure}

The high resolving power of the PEPSI spectra ($R\simeq 250,000$; corresponding to 1.2\,\kms\, or 0.024\,\AA\ at 6000\,\AA), in combination with the mean full width of the spectral lines at continuum level ($2 \ \lambda/c \ v\sin i \simeq 0.6$\,\AA), provides 25 resolution elements across the projected stellar disk. Following the simulations by \citet{pis:weh}, which suggest that at least five resolution elements are required for successful DI, the PEPSI data are properly dimensioned for surface reconstruction. In contrast, the lower resolving power of STELLA+SES spectra ($R$=55,000) renders them a border case for DI in this case.

The computation of local line profiles within $i$MAP involves solving the radiative transfer equation for each surface pixel across 72 depth points, using a grid of tabulated Kurucz ATLAS-9 model atmospheres \citep{Kurucz1993, atlas9}. The local line profiles are calculated under the assumption of one-dimensional (1D) geometry and local thermodynamic equilibrium (LTE). The atmospheric grid spans effective temperatures from 3500\,K to 8000\,K in increments of 250\,K, interpolated to match the stellar surface gravity, metallicity, and microturbulence parameters obtained from global spectrum synthesis. The stellar surface is divided into  5$^\circ$$\times$5$^\circ$ elements, corresponding to a grid of 72$\times$36 pixels and totaling 2592 surface segments. A stellar inclination of $56^\circ$, which is calculated by using the stellar parameters obtained from the literature and ParSES results. The surface reconstruction employs an iteratively regularized Landweber method, implementing a fixed-point iteration scheme designed to minimize the sum of the squared residuals. The method takes the noise level of the data set as a fixed-point and stops the iteration when the sum of the squared residuals reaches the noise limit. In other words, this iteration technique is in a way in the opposite direction when compared with other methods such as Tikhonov or Maximum Entropy, where the noise level is used as an additional term to the $\chi^{2}$ while applying the iteration \citep[see][and references therein]{carroll12}. Our final image has been achieved with a $\chi^{2}$ of 0.005.

\subsection{Adopted stellar parameters} \label{Sect_St_Pars}

As we disentangled both spectra, stellar parameters are obtained for both stars. For the photospheric parameters, we compared the SES spectra with synthetic templates. Our code Parameters-from-SES (ParSES; \citet{all04, parses}) is a software package which employs a grid of precomputed model spectra generated with Turbospectrum \citep{turbo} and a line list obtained from VALD-3 \citet{vald3}. ParSES selects the best-fit to the observed spectra based on the minimum distance method with a non-linear-simplex optimization \citep{all}. We applied this procedure to 122 STELLA spectra observed during a time coverage of 260 days for the primary star. ParSES provided in total five stellar parameters: effective temperature $T_{\rm eff}$, projected rotational velocity $v\sin i$, microturbulence $\xi_{\rm t}$, gravity $\log g$, and metallicity [M/H]. The calculated values for those parameters are as follows; $T_{\rm eff}$=4480$\pm$70\,K, $v\sin i$=15.4$\pm$1.2\,\kms, $\xi_{\rm t}$=2.0$\pm$0.2\,\kms, $\log g$=2.8$\pm$0.2, and [M/H]=$-$0.19$\pm$0.10. Errors are assumed from the 1$\sigma$ rms values from the 122 individual spectra. Our $v\sin i$ and $\xi_t$ values agree with the one previously computed by \citet{Osten1998}. \cite{Strassmeier1989} showed that the visual magnitude change of HR\,7275 has a varying amplitude from 0.002 to 0.2\,mag. This change already affects the determination of $T_{\rm eff}$ by $\simeq$230\,K. At least for the time range of our observations, those values are similar to the values calculated by \citet{O'neal1996} and \citet{Osten1998}. By using the parallax from \citet{DR3}, we calculate a luminosity of 30.6$\pm$2.8\,$L_{\odot}$\,based on $V$=5.893\,mag, $A_{\rm V}=0$, and a bolometric correction of $-$0.48 from \citet{Popper1980}. The Stefan-Boltzmann law then suggests a likely radius of 9.2\,$R_{\odot}$ (primary component). Combining the ParSES value for $\log g$ and this radius gives a likely mass for the primary star of 1.93$\pm$0.45\,$M_{\odot}$. By using this formal mass, we estimate a most-likely inclination of the orbital plane of $\approx$52$\degr$. 

Prior to the iterative Doppler-imaging process, we run several test solutions with different inclinations of the rotational axis $i$ (of the primary). Solutions in five-degree steps from $i=20\degr$ to $i=70\degr$ and with a fixed $v\sin i$ provide a best fit at $i=50\degr$ as judged from the minimum of the achieved $\chi^2$ with an equally likely range of approximately $\pm$7$\degr$. This value agrees very well with the above determination of the inclination of the system. If we assume that the rotational axis of the primary is perpendicular to the orbital plane, then we can use this inclination also for the orbital elements and obtain real masses.

\begin{table}
  \caption{Stellar parameters obtained in this paper.}\label{ParSES_Table}
\begin{tabular}{l r r r r c}
\hline\hline\noalign{\smallskip}
Parameters & Primary & Secondary  \\
\noalign{\smallskip}\hline\noalign{\smallskip}
$T_{\rm eff}$ \ [K]  & 4480$\pm$70 & 5530$\pm$116 \\
$\log g$ \ (cgs) & 2.8$\pm$0.3 & 4.0$\pm$0.2 \\
$v\sin i$ \ [\kms] &  15.4$\pm$1.2 & 1.4$\pm$0.3 \\
$\xi_{\rm t}$ \ [\kms] & 2.0$\pm$0.2 & 0.8$\pm$0.2 \\
$\rm [M/H]$ &  $-$0.19$\pm$0.10 & $-$0.18$\pm$0.11 \\
$P_{\rm rot}$ \ [d]      & = $P_{\rm orb}$ & \dots \\
$i$ \ [$\degr$]         & 52$\pm$8 & \dots \\
$M$ \ [$M_{\odot}$]         & 1.93$\pm$0.45 & 1.75$\pm$0.48 \\
A(Li)       &0.58$\pm$0.1  &0.16$^{+0.23}_{-0.63}$\\ 
MK class$^1$            & K2\,IV-III & G4\,V-IV\\
\noalign{\smallskip}\hline
\end{tabular}
\tablefoot{$^1$Based on above $\log g$ and $T_{\rm eff}$ and the tables in \citet{Gray2022}.}
\end{table}

\begin{table}
  \caption{Spots on HR\,7275a in May-June 2022.}\label{DI_spots}
\begin{tabular}{l r r r r c}
\hline\hline\noalign{\smallskip}
Spot  & Long & Lat & $\Delta T_{\rm spot}$ & Area \\
ID    & (\degr)& (\degr) & (K) & (\% ) \\
\noalign{\smallskip}\hline\noalign{\smallskip}
A (umbra)       & 315 &  25  & 1000 &  14 \\
A (penumbra)    &     &       &  350 &  6 \\
B (umbra)       & 235 &  35  & 900 &  8 \\
B (penumbra)    &     &       & 300   &  7 \\
C     & 185 &20   & 500  &  14 \\
D     & 135 &  30 & 500  &  12  \\
E (umbra)       & 60 &  35  & 1000 &  13 \\
E (penumbra)    &     &       &  450 &  6 \\
\noalign{\smallskip}\hline
\end{tabular}
\tablefoot{Longitudes and latitudes are given for the spot center. The spot temperature difference is given for the coolest part of the spot. Three spots show possibly umbral and penumbral features (A, B and E). The spot area is given in per cent of the visible hemisphere.}
\end{table}

Because our Doppler imaging assumes a perfectly spherical star, we have estimated the oblateness of the primary due to tidal forces or rapid rotation based on our orbit in Table~\ref{hr7275_orb}, the stellar parameters in Table~\ref{ParSES_Table}, and a range of Love numbers following \citet{Leconte2011}. The range for the expected polar-to-point radii never exceeded 0.38--1.4\,\%.  This verifies that our spherical-shape assumption is a reasonably good representation as the star does not experience  significant elongation. 

We also calculate stellar parameters for the secondary star using ParSES. For this, we employ a disentangled PEPSI spectrum of the secondary star (shown in the Appendix in Fig.\,\ref{hr7275b_spec}) obtained via the median-subtraction method explained above. The stellar parameters for the secondary star were obtained as follows; $T_{\rm eff}$=5530$\pm$116\,K, $v\sin i$=1.4$\pm$0.3\,\kms, $\xi_{\rm t}$=0.8$\pm$0.2\,\kms, $\log g$=4.0$\pm$0.2, and [M/H]=$-$0.18$\pm$0.11. The stellar parameters for both stars are summarized in Table\,\ref{ParSES_Table}. Because we have an SB2 orbital solution, we find the secondary mass to be 1.75$\pm$0.48\,$M_{\odot}$.

\subsection{Doppler imaging results}

The spot configuration of HR\,7275a at the time of our observations in May-June 2022 is shown in Fig.\,\ref{DI_HR7275}. The image has been obtained from a single rotation of the star and is presented via an orthographic projection with four equally-spaced rotational phases. Figure\,\ref{Mercator_DI} shows the same image in a pseudo-Mercator projection for a more global view of the stellar surface. We did not reconstruct a cool (dark) polar spot. Instead, we see a weakly warmer (bright) region about 100\,K warmer than the quiet surface of HR\,7275a. It is likely unreal because the expected temperature uncertainty in the reconstruction is about 100\,K, mostly due to the (comparably) low rotational line broadening but also the remaining uncertainties due to our LTE assumption. 

We observed in total five cool spots that appear  separated in longitude on average $60^{\circ}$ from each other. The spot locations, temperature differences, and surface coverages compared with the visible stellar surface are quantified and given in Table\,\ref{DI_spots}. Among those five spots, three (denoted A, B and E) show possibly a solar-like umbral and penumbral structure with an umbral spot temperature difference $\Delta T$ of $\simeq$1000\,K cooler than the quiet stellar photosphere. The spots named C and D appear significantly warmer than spots A, B, and E, with $\Delta T$ of $\simeq$500\,K relative to the photosphere. Both are located close-together in the same region on the stellar surface at the phase of quadrature at $\approx$0.6, that is, in the trailing hemisphere $90^{\circ}$ behind the binary's apsidal line. 

\begin{table}
\caption{Logarithmic absolute emission-line fluxes in \ergs\ for HR\,7275.}\label{T_fluxes}
\begin{tabular}{lllll}
\hline\hline\noalign{\smallskip}
Bandpass  & Continuum       & Average    & Average       & Fit   \\
          & flux            & line flux  & variability   & rms   \\
          & (per \AA )      &            & amplitude     &       \\
\noalign{\smallskip}\hline\noalign{\smallskip}
Ca\,{\sc ii} H    & 6.2060 & 6.135 & 0.571   & 0.056\\
Ca\,{\sc ii} K    & 6.2060 & 6.247 & 0.732   & 0.084\\
\Halpha           & 6.4462 & 6.026 & 0.296   & 0.033\\
Ca\,{\sc ii} 8542 & 6.3911 & 6.154 & 0.141   & 0.013\\
\noalign{\smallskip}\hline
\end{tabular}
\tablefoot{Variability amplitude and fit rms of relative fluxes in units of the continuum ($=1$)}
\end{table}

\section{Magnetic activity and lithium abundance} \label{Chromo_Sect}

\subsection{\cahk, \Halpha\ and \cairt\,8542\,\AA\ line-core emissions}

Four chromospheric activity indicators are being used in this paper: \cahk, \Halpha, and one of the infrared triplet lines \cairt\ (8542\,\AA). As shown in Fig.\,\ref{Emissions}, these diagnostics are available for a total of six rotations; four consecutive stellar rotations are within the $BJD$ 2,459,653-766 window, while another two consecutive rotations are tracked between 2,459,824 and 2,459,882.

\begin{figure}[t]
    \includegraphics[width=87mm]{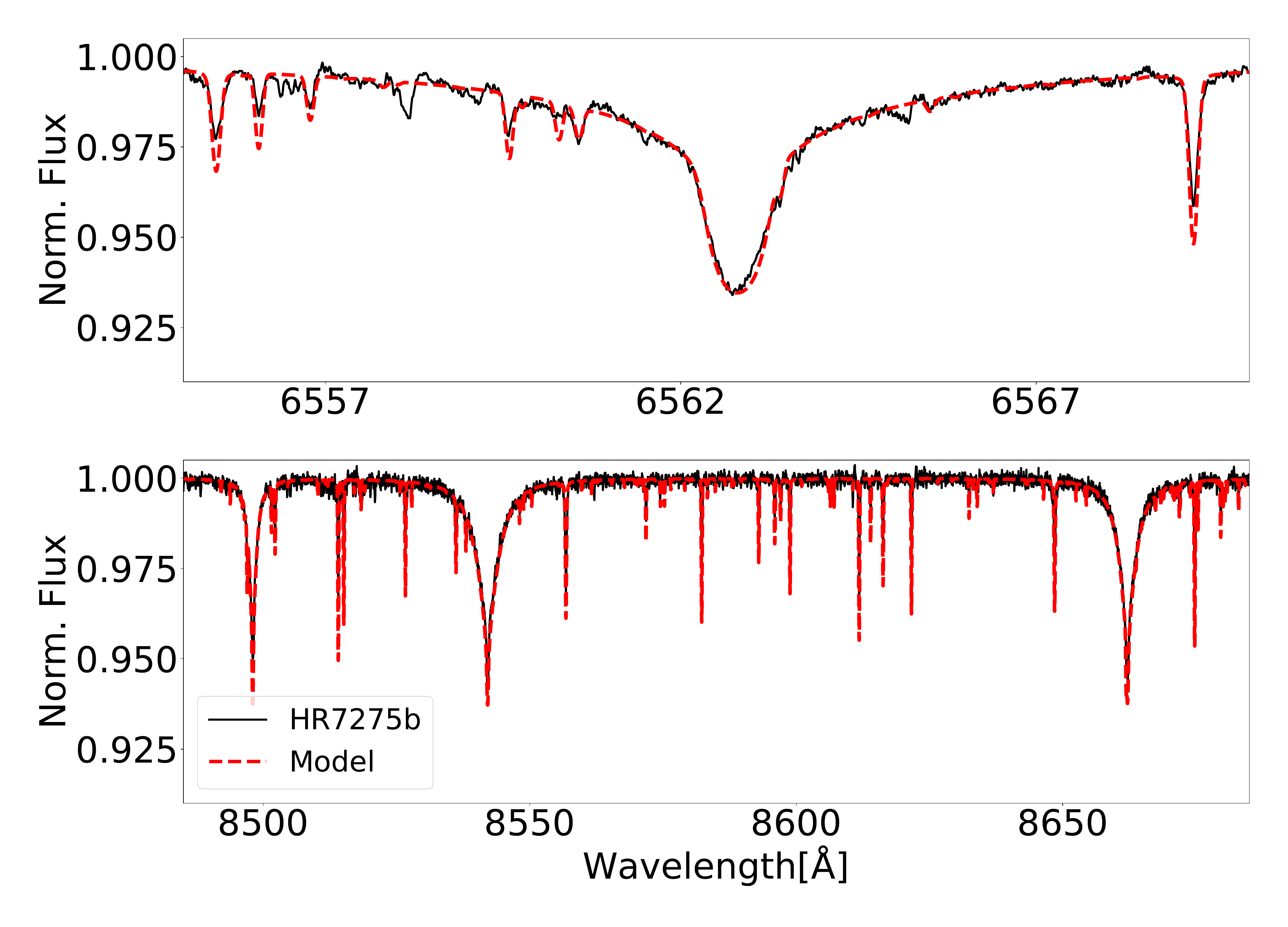}
    \caption{Disentangled spectrum of HR7275b near \Halpha\,(upper panel) and \cairt\ region (lower panel). The observed spectrum is indicated by a black line and the model is indicated by dashed red line.}
    \label{Halpha_cairt_starB}
\end{figure}

We measure the relative flux in a bandwidth of one \AA\ centered on the line core of 3968.5\,\AA\ and 3933.7\,$\AA$ for \cahk, 8542.1\,$\AA$ for \cairt\ and 6562.8\,$\AA$ for \Halpha. For the latter two lines, we have spectra from both STELLA and PEPSI while the \cahk\ analysis is done only with STELLA data (PEPSI+VATT spectra did not cover the bluest wavelength regions due to the long fiber). The surface temperature calculations are based on ParSES as explained in Sect.\,\ref{Sect_St_Pars}. The fluxes for \Halpha\ and \cairt\ are computed after the disentangling for the corresponding wavelength regions. The obtained secondary star spectrum for \Halpha\,and \cairt\,regions are shown in Fig.\,\ref{Halpha_cairt_starB}. Because our disentangling processes fail for the \cahk region mostly because of too low S/N and the intrinsic variability of the primary, we calculated \cahk\ line fluxes from the composite spectra. We note that the relative contribution of the secondary must be rather small at this wavelength region. The reason for this is that the $v\sin i$ of the secondary component is very small, thus the secondary star is not expected to be very active. In addition, as a nature of the RS\,CVn systems, the strong line emissions of the primary star are expected to dilute the signatures coming from the companion.

\begin{figure}[t]
{\bf a.} \\
    \includegraphics[width=87mm]{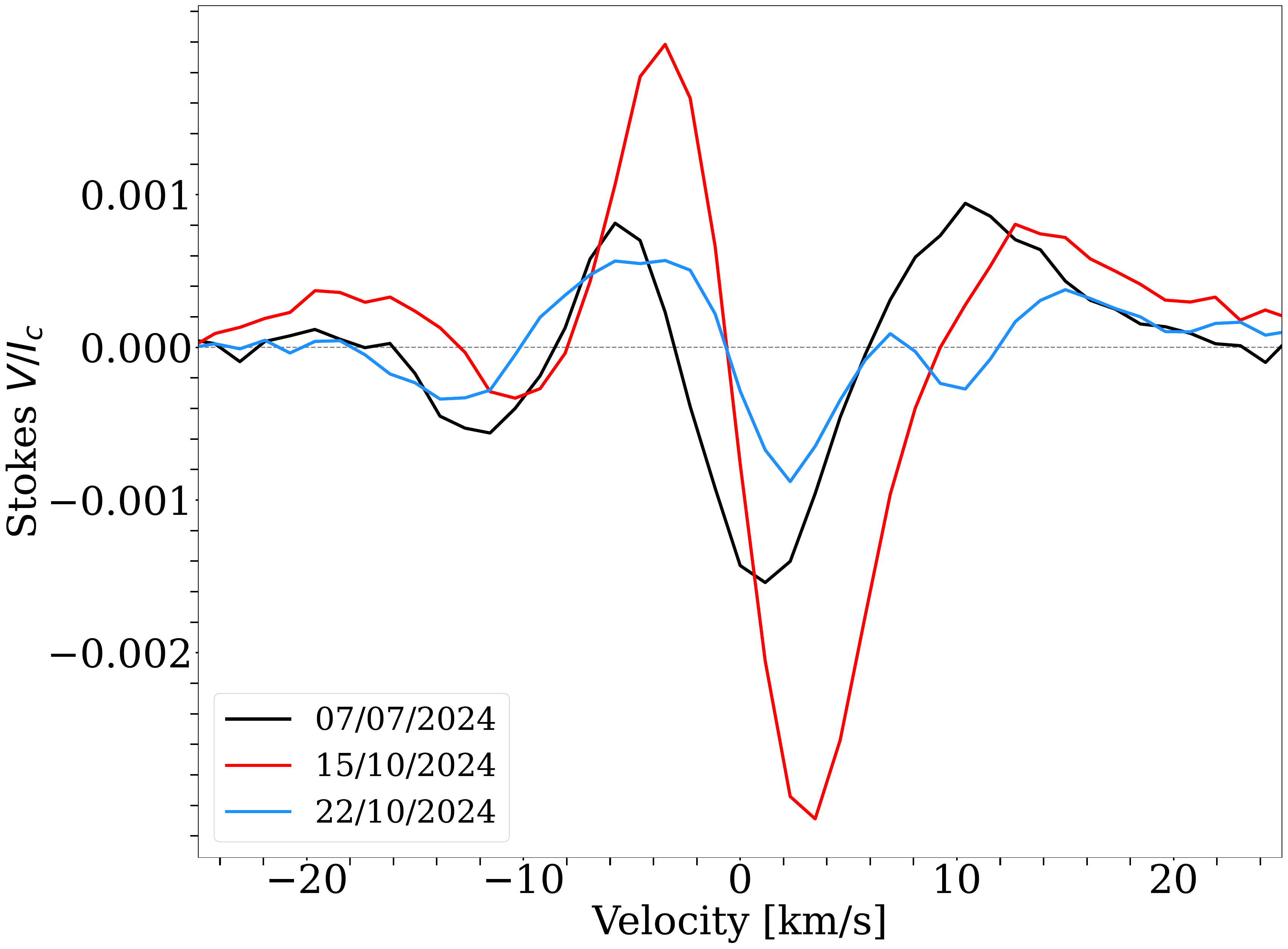}

{\bf b.} \\
    \includegraphics[width=87mm]{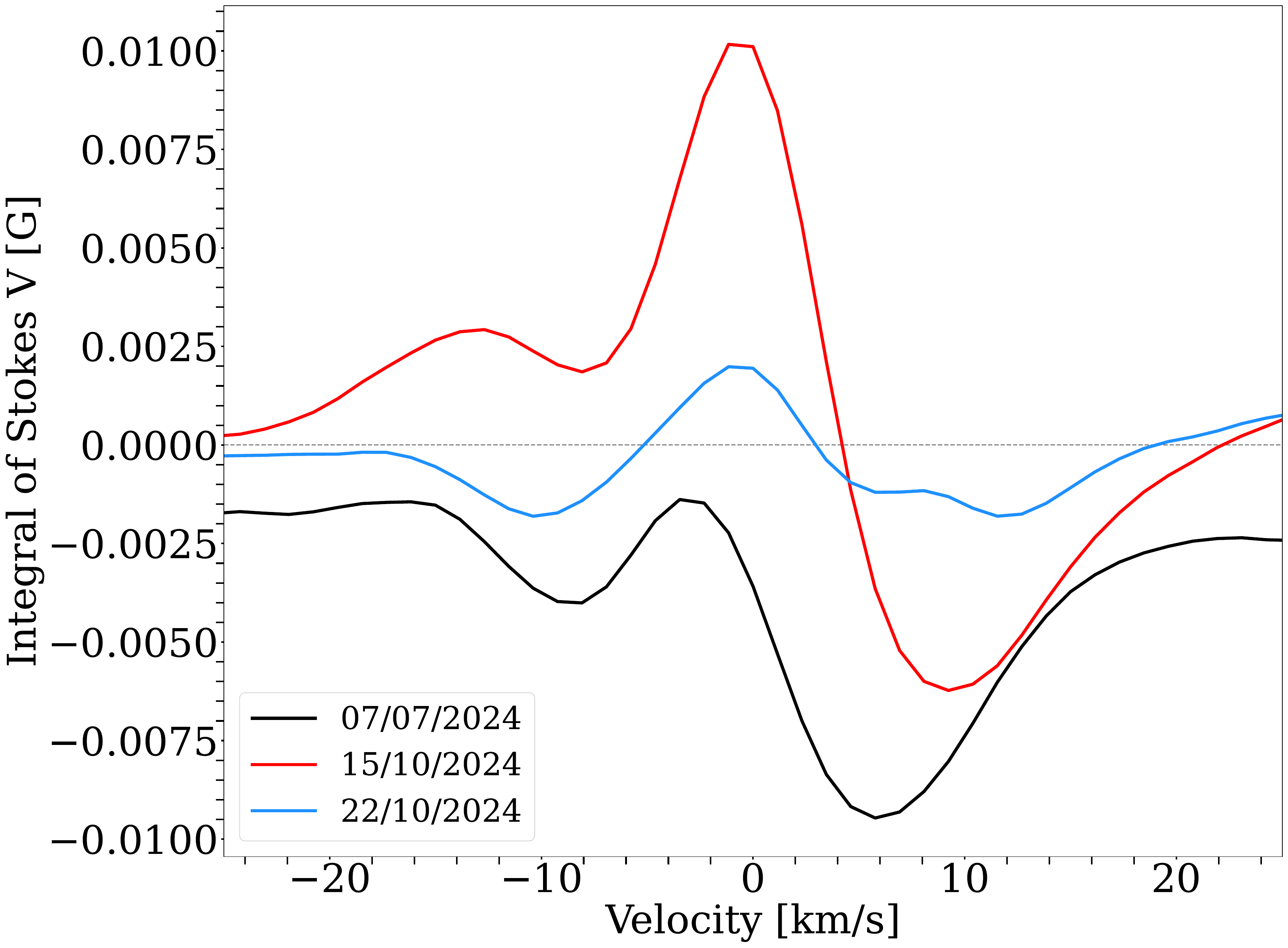}
    \caption{Stokes-V based magnetic field measurements for HR\,7275 from 2024. \emph{a.} LSD line profiles from PEPSI in units of normalized intensity. \emph{b.} Integrated Stokes-V profiles in units of Gauss. }
    \label{StokesV}
\end{figure}

Before the conversion to absolute flux, we match the continuum settings of STELLA spectra with that from PEPSI spectra because of the latter superiority in quality and data reduction \citep[see][]{Jarvinen2025}. Intrinsic flux-variability amplitudes are determined by three sinusoidal fits for each of the four consecutive stellar rotation. The average amplitude of the chromospheric (flux) variability and its rms values with respect to these fits are given in Table\,\ref{T_fluxes}.

The chromospheric activity of HR\,7275 appears strongly modulated with either the rotational period ($\simeq$28\,d) or half of it ($\simeq$14\,d) depending upon its single- or a double-humped appearance (see Table\,\ref{Period_Analysis}). As shown in the Appendix in Fig.\,\ref{Emissions}, the maximum amplitude change is about 35\% for \cahk\ and \cairt\ 8542\,\AA, while it is 40\% for \Halpha. The peak-amplitude location of these emissions appears at shifts of about 0.1 phases  per rotation, despite that the overall activity modulation appears stable during the first four rotations. During the third of the covered six full stellar rotations, that is the time of our photospheric Doppler imaging, we observe a complex double-peaked flux distribution with the strongest peak at approximately phase 0.55 (Fig.\,\ref{DI_HR7275}c), which coincides with the position of the two weakest (warmest) spots D and C during their central-meridian passage. The two-times weaker secondary peak appears around phase 0.15 and coincides with the central-meridian passage of the large spot~A. Both peaks  appear similarly in the other three chromospheric tracers. The main peak at phase 0.55 is particularly prominent in \Halpha\ which is indicative of a facular origin, or at least contribution, compared to solar analogy. 

\subsection{Magnetic field measurements}

We obtained magnetic field measurements from three PEPSI Stokes-V spectra in 2024. The spectra cover the wavelength range from 4800\,\AA\ to 5441\,\AA\ and are converted to Least Squares Deconvolution \citep[LSD; e.g.][]{Donati1997} line profiles built from the most recent line list from VALD-3 \citep{vald3}. For the magnetic field measurement, we implemented the prescription from \citet{Kochukhov2010} based on the method of \citet{Sol&Ste1984}. The three LSD profiles are shown in Fig.\,\ref{StokesV}a. 

\begin{table}
\caption{Magnetic field measurements for HR\,7275.}\label{T_magfield}
\begin{tabular}{llllll}
\hline\hline\noalign{\smallskip}
BJD & Phase & \multicolumn{2}{c}{$B_{\rm long}$}      &|$B$|   \\
    & $(\varphi)$ & \multicolumn{2}{c}{[G]}                & [G]     \\ 
\noalign{\smallskip}\hline\noalign{\smallskip}
2,460,498.973 & 0.611 & \multicolumn{2}{c}{$-15.2 \pm 2.7$}   & \multicolumn{2}{c}{$57.1 \pm 0.7$}   \\
2,460,598.588 & 0.097 & \multicolumn{2}{c}{$+0.6  \pm 2.0$}   & \multicolumn{2}{c}{$35.2 \pm 0  .7$}  \\
2,460,605.686 & 0.345 & \multicolumn{2}{c}{$-14.0 \pm 2.6$}   & \multicolumn{2}{c}{$17.8 \pm 0.6$}  \\
\noalign{\smallskip}\hline
\end{tabular}
\end{table}

The disk-integrated longitudinal magnetic field was $-$15.2$\pm$2.7\,G for the one spectrum from July 2024. For the two observations in October 2024, seven days apart, we obtained rather divergent results of +0.6$\pm$2.0\,G and $-$14.0$\pm$2.6\,G. While we can not conclude on the global morphology of the field from these three measurements, it indicates the presence and phase dominance of even and odd polarities coexisting within short (phase) distances on the surface.   

\subsection{Lithium abundance}

In order to get an estimate for the mixing inside the star, we determine the logarithmic lithium abundance, A(Li), of the primary and also attempt to measure it for the secondary star. After disentangling the spectra within the region between 6300-7400\,\AA, using the median subtraction technique explained in Sect.\,\ref{Med_Subt}, we first built a median spectrum from all 26 PEPSI spectra of the primary and the secondary star. These two steps, disentangling and phase averaging, enabled not only the removal of the respective binary spectrum but also identified and removed the telluric contamination. This single spectrum is of high quality with a median S/N per pixel of 300:1 for the primary and 99:1 for the secondary. 

\begin{figure}[t]
{\bf a.} \\
    \includegraphics[width=87mm]{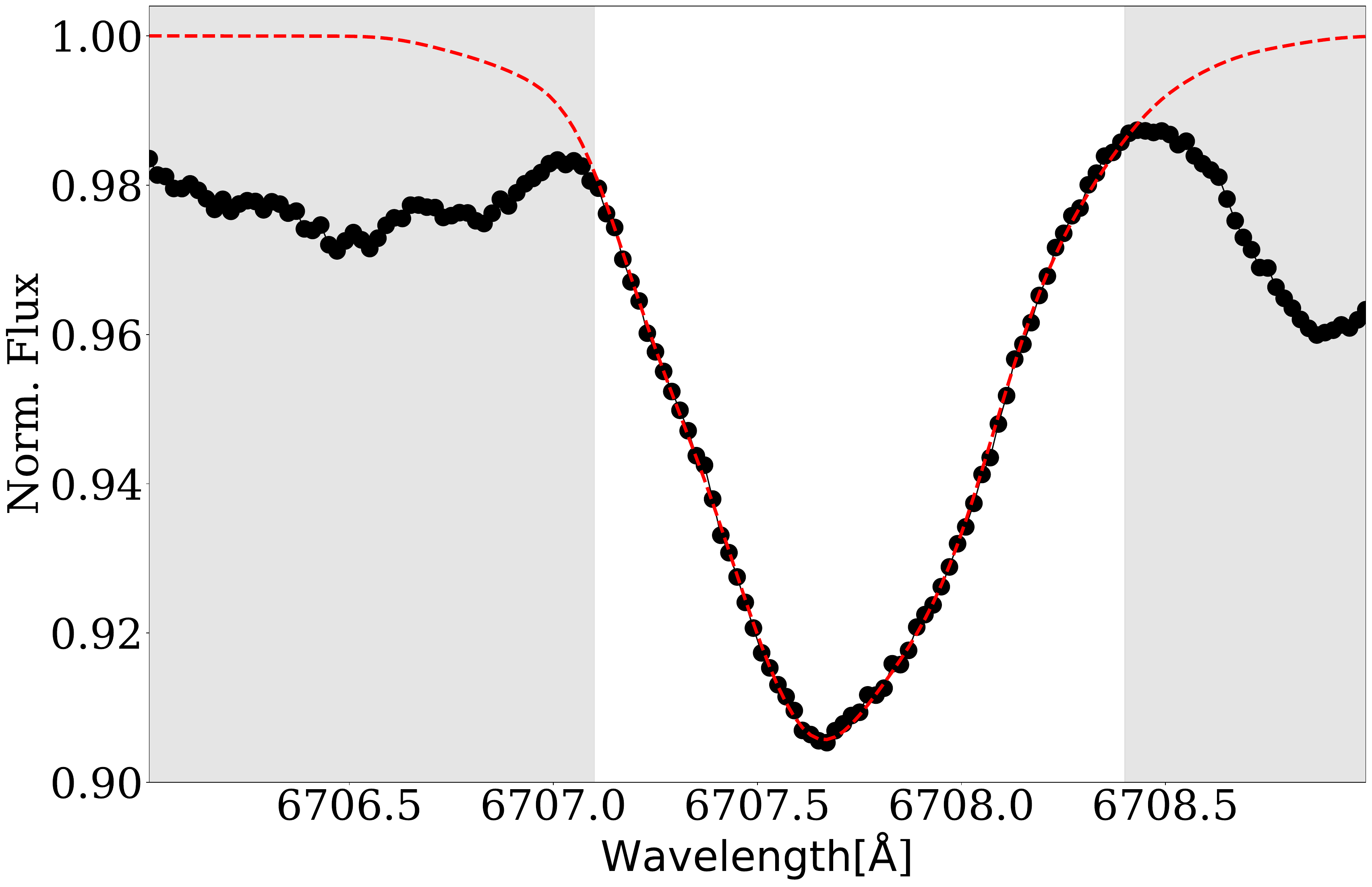}

{\bf b.} \\
    \includegraphics[width=87mm]{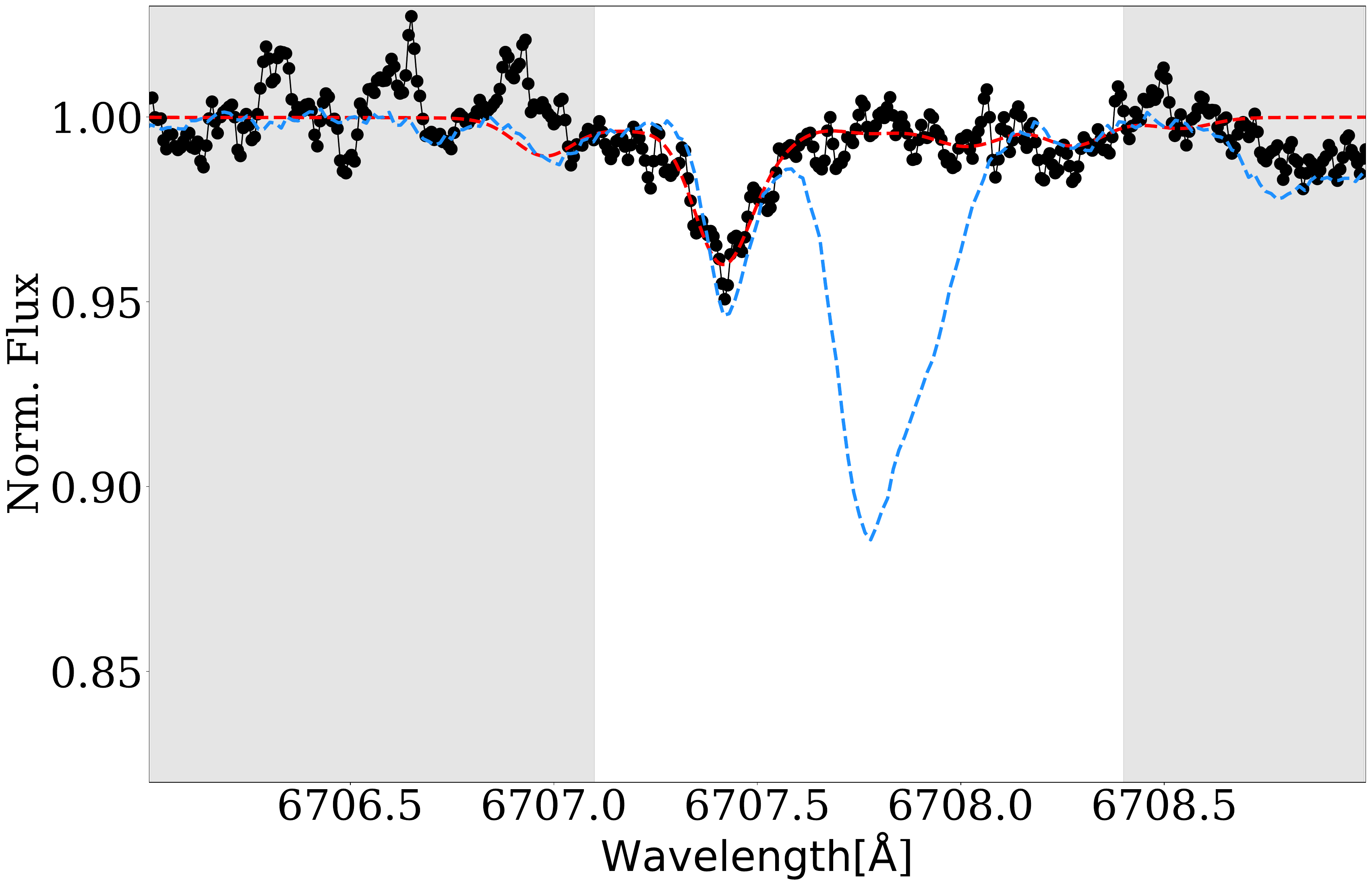}
    \caption{Lithium abundance analysis. Panel \emph{a}: Median spectrum of HR\,7275a (black dots) and Turbospectrum fit (dashed red line). The gray shaded areas indicate the regions outside the fitting range (in white). Panel \emph{b}: Same plot for HR\,7275b. The blue dashed line shows a PEPSI spectrum of 70\,Vir. }
    \label{Lithium_fit}
\end{figure}

We then employed the line-synthesis program Turbospectrum \citep{turbo} together with the fitting routine TurboMPfit \citep{Steffen2015} and the line list from \citet{Melendez2012}. Many different fits were obtained for the primary by both including and excluding TiO lines, fixed and adapted continuum level, line shifts, and macroturbulence broadening. Somewhat surprisingly, we found that the impact of TiO molecular lines is very moderate despite the low $T_{\rm eff}$ of 4480\,K. Presumably, this can be explained by the lower particle densities in the line formation regions of this giant star at $\log g$=2.8. Figure\,\ref{Lithium_fit}a shows the best fit with an abundance of 0.58$\pm$0.1, with or without TiO blends. The comparably large error, $\pm$0.1\,dex, is by far due to the $\pm$70\,K uncertainty of the ParSES-based $T_{\rm eff}$ rather than anything else. The internal fit quality of all our Turbospectrum runs was always around or even below 0.01\,dex, thus a factor ten lower, as the fit is so much more sensitive to the overall photospheric effective temperature. In addition, the contribution of TiO lines does not affect the fit significantly, both versions result in abundances less than 1\% different. Although HR\,7275a displays spots cooler than the surrounding photosphere by 1000\,K, a fact which should favor local TiO line formation, the effective temperature of the star is just too high to allow the detection of these lines \citep{Strassmeier2022}. Similar observations and conclusion were reported by \citet{Adebali2025} for \lama, which is only about 100\,K hotter than HR\,7275a. 

A similar procedure than above was applied to the secondary spectrum with the difference that we adopted the continuum at 6708\,\AA\ as a free parameter from the beginning on and fixed line shift and total line broadening. This left only the continuum and the Li abundance as free parameters. The reason is simply the comparable low S/N. Fits were obtained for three effective temperatures, $\pm$115\,K around the nominal $T_{\rm eff}$ of 5530\,K, and three relative metallicities 0, $-0.1$, and $-0.2$. The best fit was obtained for $T_{\rm eff}$=5530\,K and [Fe/H]=$-0.1$ with $A$(Li)=0.16\,dex and is shown in Fig.~\ref{Lithium_fit}b. Its internal error is more or less meaningless but we estimate a real error from the range of $\pm$115-K in temperature and $\pm$0.1\,dex in metallicity, which then suggests only a marginal detection of lithium of $A$(Li)=+0.16$^{+0.23}_{-0.63}$.     
The final result for the lithium abundance of the primary is very robust, $A$(Li)=0.58$\pm$0.1\,dex, indicating considerable mixing as expected for an evolved star. The $^6$Li/$^7$Li isotopic ratio cannot be deduced from the highly broadened spectrum and was fixed at 0.0. The lithium abundance of the secondary  seems to be extremely low. For comparison, $A$(Li)=1.12 for the single benchmark star 70\,Vir (also classified as G4\,V-IV), so roughly a factor 10 higher than for HR7275b.

\section{Summary and conclusions}\label{Summary}

In this paper, we used different techniques to understand the activity behavior of the primary star of the system HR 7275. We developed a \texttt{python} tool called \texttt{DISTRACT} using two different methods for isolating the spectra of two stars. This enabled us to quantify the following activity indicators for HR 7275a.

The average RV jitter of HR\,7275a is comparable to what has been observed for the highly asynchronous primary star in the \lama\ system \citep{Adebali2025}. The RV contribution from HR\,7275a spots have two different global modulation behavior. The first modulation pattern causes up to 320\,\ms\ of RV jitter and the second global modulation peaks at 650\,\ms. Our global fits for these modulations gives an estimate for the general activity behavior of the star. Nevertheless, the activity modulation differs further for each rotation of the star. Between BJD\,2,459,640-2,459,828, we had four consecutive rotations with good enough sampling rate. During the first rotation, the RV modulation reaches up to 600\,\ms\ but diminishes to 200\,\ms\ in the next two rotations resulting in an average of 320\,\ms\ when including the other two rotations. After BJD\,2,459,828, we had two more but less-well sampled rotations. Both shows a larger spread in RV jitter than before and a higher jitter average of 650\,\ms. The reason for this change is simply the disk asymmetry caused by the spot distribution with respect to each other and the stellar limb. This may be a feature that separates HR\,7275a from its asynchronously rotating RS\,CVn cousin \lama, which shows a stable spot coverage, and thus stable RV jitter, during 522\,d of observations (almost ten stellar rotations). In addition, the occurrence of three relatively large spots, each occupying almost 10\%\ of the surface of HR\,7275a, is indicative that the star has a more intense dynamo process than \lama. At this point, we must conclude that synchronized rotation does not necessarily favor more stable magnetic structures, and therefore a more stable stellar magnetic activity.

This aspect is also observed by chromospheric emissions. As shown in Fig.\,\ref{Emissions}, the activity peaks appear at different relative flux values. For \cahk\ and \cairt\,8542 emissions, the highest activity appears during the second stellar rotation. Contrary to that, \Halpha\ emission flux peaks at the third rotation where we also construct the Doppler image of the primary star. This shows that, emission fluxes in \Halpha\ regime do no exactly follow the \cairt\ and \cahk. When we consider the formation height of those emission lines, we conclude that the more intense dynamo behavior affects different layers of the chromosphere with a varying impact.

Characterizing the activity jitter on RVs and confirming these effects with Doppler image gives a broader understanding for how starspot structure affects the chromospheric emissions. As shown in Fig.\,\ref{DI_HR7275}, relatively warmer spots cause a larger chromospheric emission on \cahk. We interpret this as a faculae dominated region heats up the upper atmosphere more than we see above the cooler spots on HR\,7275a. This effect is also observable in the Sun, where magnetic field is the main source of the heating mechanism in the chromosphere. The dominating spots marked with C and D appear also closer to each other than the other resolved spots. This may also enhance the emission more than the observed single spots (A and E), while they are passing from the central meridian.

Monitoring the surface activity allowed us to  determined both SB1 and SB2 solutions for the system by correcting the activity ``jitter''. These corrections allowed us to improve the orbital solution by 35\% for the SB1 calculation and by about 5\% for the SB2 calculation. The reason for the latter being relatively low is the fact that we cannot detect surface effects on RVs for the secondary star likely because of its very low $v\sin i$ and the low relative activity signal appearing in the composite spectra. In the near future, we plan for a simultaneous RV and activity fit with different analysis techniques such as Gaussian processes \citep[e.g.,][]{Aigrain2023}.

The improved orbit also enabled more accurate stellar parameters for both primary and secondary star. The most recent orbital calculations in the literature by \citet{Medeiros1999}, suggested similar minimum masses for the primary as listed in Table\,\ref{hr7275_orb}. However, since there is no information provided about the inclination by \citet{Medeiros1999}, they could  not provide values for the actual masses. Our best value for the primary mass is 1.93$\pm$0.45\,M$_\odot$ based on an inclination of 52$\pm$8$\degr$ from ParSES calculations. \citet{Eker1989} suggested a mass range for the secondary star between 0.9-1.1\,M$_{\odot}$ by using their SB1 solution. When we combine our SB2 solution with the Stefan-Boltzmann equations, we found a consistent but much higher value of 1.75\,$M_{\odot}$. 

Finally, we determined lithium abundances for both components of the system for the first time. The primary star has $A$(Li)=0.58\,dex and the secondary star of $A$(Li)$\approx$0.16. The latter being only a marginal detection. The comparably low abundance of the giant primary indicates that its photospheric abundance must have undergone significant mixing.

\begin{acknowledgements}
We thank an anonymous referee for the constructive and detailed comments that improved the quality of this article. This work is based partially on data obtained with the Stellar Activity-2 (STELLA-II) robotic telescope in Tenerife, an AIP facility jointly operated by AIP and IAC (https://stella.aip.de/) and partially on data from PEPSI acquired with the Large Binocular Telescope (LBT) and the Vatican Advanced Technology Telescope (VATT) (see https://pepsi.aip.de/). The LBT is an international collaboration among institutions in the United States, Italy and Germany. LBT Corporation partners are: The University of Arizona on behalf of the Arizona Board of Regents; Istituto Nazionale di Astrofisica, Italy; LBT Beteiligungsgesellschaft, Germany, representing the Max-Planck Society, The Leibniz Institute for Astrophysics Potsdam, and Heidelberg University; The Ohio State University, representing OSU, University of Notre Dame, University of Minnesota and University of Virginia.
In this work, we heavily used \texttt{python3} libraries; \texttt{astropy}\, \citep{astropy:2013, astropy:2018, astropy:2022},\, \texttt{numpy}\, \citep{numpyharris2020}\, and \texttt{scipy}\, \citep{SciPy2020-NMeth}.
The authors thank B. Seli from Konkoly Observatory for making available his Python code for spot segmentation \citep[see ][Appendix F]{Kovari2024}. 
ZsK acknowledges the financial support of the Hungarian National Research, Development and Innovation Office grant KKP-143986.

\end{acknowledgements}

\bibliography{ozgun} 

@ARTICLE{Adebali2025,
       author = {{Adebali}, {\"O}. and {Strassmeier}, K.~G. and {Ilyin}, I.~V. and {Weber}, M. and {Gruner}, D. and {K{\H{o}}v{\'a}ri}, Zs.},
        title = "{First Doppler image and starspot-corrected orbit for {\ensuremath{\lambda}} Andromedae: A multifaceted activity analysis}",
      journal = {\aap},
         year = 2025,
        month = mar,
       volume = {695},
          eid = {A89},
        pages = {A89},
          doi = {10.1051/0004-6361/202453073},
       adsurl = {https://ui.adsabs.harvard.edu/abs/2025A&A...695A..89A}
}

@ARTICLE{all,
       author = {{Allende Prieto}, Carlos and {Beers}, Timothy C. and {Wilhelm}, Ronald and {Newberg}, Heidi Jo and {Rockosi}, Constance M. and {Yanny}, Brian and {Lee}, Young Sun},
        title = "{A Spectroscopic Study of the Ancient Milky Way: F- and G-Type Stars in the Third Data Release of the Sloan Digital Sky Survey}",
      journal = {\apj},
         year = 2006,
        month = jan,
       volume = {636},
       number = {2},
        pages = {804-820},
          doi = {10.1086/498131},
archivePrefix = {arXiv},
       eprint = {astro-ph/0509812},
 primaryClass = {astro-ph},
       adsurl = {https://ui.adsabs.harvard.edu/abs/2006ApJ...636..804A}
}

@ARTICLE{all04,
       author = {{Allende Prieto}, C.},
        title = "{Automated analysis of stellar spectra}",
      journal = {Astronomische Nachrichten},
         year = 2004,
        month = oct,
       volume = {325},
       number = {6},
        pages = {604-609},
          doi = {10.1002/asna.200410291},
       adsurl = {https://ui.adsabs.harvard.edu/abs/2004AN....325..604A}
}

@article{astropy:2013,
Adsurl = {http://adsabs.harvard.edu/abs/2013A%26A...558A..33A},
Archiveprefix = {arXiv},
Author = {{Astropy Collaboration} and {Robitaille}, T.~P. and {Tollerud}, E.~J. and {Greenfield}, P. and {Droettboom}, M. and {Bray}, E. and {Aldcroft}, T. and {Davis}, M. and {Ginsburg}, A. and {Price-Whelan}, A.~M. and {Kerzendorf}, W.~E. and {Conley}, A. and {Crighton}, N. and {Barbary}, K. and {Muna}, D. and {Ferguson}, H. and {Grollier}, F. and {Parikh}, M.~M. and {Nair}, P.~H. and {Unther}, H.~M. and {Deil}, C. and {Woillez}, J. and {Conseil}, S. and {Kramer}, R. and {Turner}, J.~E.~H. and {Singer}, L. and {Fox}, R. and {Weaver}, B.~A. and {Zabalza}, V. and {Edwards}, Z.~I. and {Azalee Bostroem}, K. and {Burke}, D.~J. and {Casey}, A.~R. and {Crawford}, S.~M. and {Dencheva}, N. and {Ely}, J. and {Jenness}, T. and {Labrie}, K. and {Lim}, P.~L. and {Pierfederici}, F. and {Pontzen}, A. and {Ptak}, A. and {Refsdal}, B. and {Servillat}, M. and {Streicher}, O.},
Doi = {10.1051/0004-6361/201322068},
Eid = {A33},
Eprint = {1307.6212},
Journal = {\aap},
Month = oct,
Pages = {A33},
Primaryclass = {astro-ph.IM},
Title = {{Astropy: A community Python package for astronomy}},
Volume = 558,
Year = 2013
}

@ARTICLE{astropy:2018,
       author = {{Astropy Collaboration} and {Price-Whelan}, A.~M. and
         {Sip{\H{o}}cz}, B.~M. and {G{\"u}nther}, H.~M. and {Lim}, P.~L. and
         {Crawford}, S.~M. and {Conseil}, S. and {Shupe}, D.~L. and
         {Craig}, M.~W. and {Dencheva}, N. and {Ginsburg}, A. and {Vand
        erPlas}, J.~T. and {Bradley}, L.~D. and {P{\'e}rez-Su{\'a}rez}, D. and
         {de Val-Borro}, M. and {Aldcroft}, T.~L. and {Cruz}, K.~L. and
         {Robitaille}, T.~P. and {Tollerud}, E.~J. and {Ardelean}, C. and
         {Babej}, T. and {Bach}, Y.~P. and {Bachetti}, M. and {Bakanov}, A.~V. and
         {Bamford}, S.~P. and {Barentsen}, G. and {Barmby}, P. and
         {Baumbach}, A. and {Berry}, K.~L. and {Biscani}, F. and {Boquien}, M. and
         {Bostroem}, K.~A. and {Bouma}, L.~G. and {Brammer}, G.~B. and
         {Bray}, E.~M. and {Breytenbach}, H. and {Buddelmeijer}, H. and
         {Burke}, D.~J. and {Calderone}, G. and {Cano Rodr{\'\i}guez}, J.~L. and
         {Cara}, M. and {Cardoso}, J.~V.~M. and {Cheedella}, S. and {Copin}, Y. and
         {Corrales}, L. and {Crichton}, D. and {D'Avella}, D. and {Deil}, C. and
         {Depagne}, {\'E}. and {Dietrich}, J.~P. and {Donath}, A. and
         {Droettboom}, M. and {Earl}, N. and {Erben}, T. and {Fabbro}, S. and
         {Ferreira}, L.~A. and {Finethy}, T. and {Fox}, R.~T. and
         {Garrison}, L.~H. and {Gibbons}, S.~L.~J. and {Goldstein}, D.~A. and
         {Gommers}, R. and {Greco}, J.~P. and {Greenfield}, P. and
         {Groener}, A.~M. and {Grollier}, F. and {Hagen}, A. and {Hirst}, P. and
         {Homeier}, D. and {Horton}, A.~J. and {Hosseinzadeh}, G. and {Hu}, L. and
         {Hunkeler}, J.~S. and {Ivezi{\'c}}, {\v{Z}}. and {Jain}, A. and
         {Jenness}, T. and {Kanarek}, G. and {Kendrew}, S. and {Kern}, N.~S. and
         {Kerzendorf}, W.~E. and {Khvalko}, A. and {King}, J. and {Kirkby}, D. and
         {Kulkarni}, A.~M. and {Kumar}, A. and {Lee}, A. and {Lenz}, D. and
         {Littlefair}, S.~P. and {Ma}, Z. and {Macleod}, D.~M. and
         {Mastropietro}, M. and {McCully}, C. and {Montagnac}, S. and
         {Morris}, B.~M. and {Mueller}, M. and {Mumford}, S.~J. and {Muna}, D. and
         {Murphy}, N.~A. and {Nelson}, S. and {Nguyen}, G.~H. and
         {Ninan}, J.~P. and {N{\"o}the}, M. and {Ogaz}, S. and {Oh}, S. and
         {Parejko}, J.~K. and {Parley}, N. and {Pascual}, S. and {Patil}, R. and
         {Patil}, A.~A. and {Plunkett}, A.~L. and {Prochaska}, J.~X. and
         {Rastogi}, T. and {Reddy Janga}, V. and {Sabater}, J. and
         {Sakurikar}, P. and {Seifert}, M. and {Sherbert}, L.~E. and
         {Sherwood-Taylor}, H. and {Shih}, A.~Y. and {Sick}, J. and
         {Silbiger}, M.~T. and {Singanamalla}, S. and {Singer}, L.~P. and
         {Sladen}, P.~H. and {Sooley}, K.~A. and {Sornarajah}, S. and
         {Streicher}, O. and {Teuben}, P. and {Thomas}, S.~W. and
         {Tremblay}, G.~R. and {Turner}, J.~E.~H. and {Terr{\'o}n}, V. and
         {van Kerkwijk}, M.~H. and {de la Vega}, A. and {Watkins}, L.~L. and
         {Weaver}, B.~A. and {Whitmore}, J.~B. and {Woillez}, J. and
         {Zabalza}, V. and {Astropy Contributors}},
        title = "{The Astropy Project: Building an Open-science Project and Status of the v2.0 Core Package}",
      journal = {\aj},
         year = 2018,
        month = sep,
       volume = {156},
       number = {3},
          eid = {123},
        pages = {123},
          doi = {10.3847/1538-3881/aabc4f},
archivePrefix = {arXiv},
       eprint = {1801.02634},
 primaryClass = {astro-ph.IM},
       adsurl = {https://ui.adsabs.harvard.edu/abs/2018AJ....156..123A}
}

@ARTICLE{astropy:2022,
       author = {{Astropy Collaboration} and {Price-Whelan}, Adrian M. and {Lim}, Pey Lian and {Earl}, Nicholas and {Starkman}, Nathaniel and {Bradley}, Larry and {Shupe}, David L. and {Patil}, Aarya A. and {Corrales}, Lia and {Brasseur}, C.~E. and {N{"o}the}, Maximilian and {Donath}, Axel and {Tollerud}, Erik and {Morris}, Brett M. and {Ginsburg}, Adam and {Vaher}, Eero and {Weaver}, Benjamin A. and {Tocknell}, James and {Jamieson}, William and {van Kerkwijk}, Marten H. and {Robitaille}, Thomas P. and {Merry}, Bruce and {Bachetti}, Matteo and {G{"u}nther}, H. Moritz and {Aldcroft}, Thomas L. and {Alvarado-Montes}, Jaime A. and {Archibald}, Anne M. and {B{'o}di}, Attila and {Bapat}, Shreyas and {Barentsen}, Geert and {Baz{'a}n}, Juanjo and {Biswas}, Manish and {Boquien}, M{'e}d{'e}ric and {Burke}, D.~J. and {Cara}, Daria and {Cara}, Mihai and {Conroy}, Kyle E. and {Conseil}, Simon and {Craig}, Matthew W. and {Cross}, Robert M. and {Cruz}, Kelle L. and {D'Eugenio}, Francesco and {Dencheva}, Nadia and {Devillepoix}, Hadrien A.~R. and {Dietrich}, J{"o}rg P. and {Eigenbrot}, Arthur Davis and {Erben}, Thomas and {Ferreira}, Leonardo and {Foreman-Mackey}, Daniel and {Fox}, Ryan and {Freij}, Nabil and {Garg}, Suyog and {Geda}, Robel and {Glattly}, Lauren and {Gondhalekar}, Yash and {Gordon}, Karl D. and {Grant}, David and {Greenfield}, Perry and {Groener}, Austen M. and {Guest}, Steve and {Gurovich}, Sebastian and {Handberg}, Rasmus and {Hart}, Akeem and {Hatfield-Dodds}, Zac and {Homeier}, Derek and {Hosseinzadeh}, Griffin and {Jenness}, Tim and {Jones}, Craig K. and {Joseph}, Prajwel and {Kalmbach}, J. Bryce and {Karamehmetoglu}, Emir and {Ka{l}uszy{'n}ski}, Miko{l}aj and {Kelley}, Michael S.~P. and {Kern}, Nicholas and {Kerzendorf}, Wolfgang E. and {Koch}, Eric W. and {Kulumani}, Shankar and {Lee}, Antony and {Ly}, Chun and {Ma}, Zhiyuan and {MacBride}, Conor and {Maljaars}, Jakob M. and {Muna}, Demitri and {Murphy}, N.~A. and {Norman}, Henrik and {O'Steen}, Richard and {Oman}, Kyle A. and {Pacifici}, Camilla and {Pascual}, Sergio and {Pascual-Granado}, J. and {Patil}, Rohit R. and {Perren}, Gabriel I. and {Pickering}, Timothy E. and {Rastogi}, Tanuj and {Roulston}, Benjamin R. and {Ryan}, Daniel F. and {Rykoff}, Eli S. and {Sabater}, Jose and {Sakurikar}, Parikshit and {Salgado}, Jes{'u}s and {Sanghi}, Aniket and {Saunders}, Nicholas and {Savchenko}, Volodymyr and {Schwardt}, Ludwig and {Seifert-Eckert}, Michael and {Shih}, Albert Y. and {Jain}, Anany Shrey and {Shukla}, Gyanendra and {Sick}, Jonathan and {Simpson}, Chris and {Singanamalla}, Sudheesh and {Singer}, Leo P. and {Singhal}, Jaladh and {Sinha}, Manodeep and {Sip{H{o}}cz}, Brigitta M. and {Spitler}, Lee R. and {Stansby}, David and {Streicher}, Ole and {{{S}}umak}, Jani and {Swinbank}, John D. and {Taranu}, Dan S. and {Tewary}, Nikita and {Tremblay}, Grant R. and {Val-Borro}, Miguel de and {Van Kooten}, Samuel J. and {Vasovi{'c}}, Zlatan and {Verma}, Shresth and {de Miranda Cardoso}, Jos{'e} Vin{'i}cius and {Williams}, Peter K.~G. and {Wilson}, Tom J. and {Winkel}, Benjamin and {Wood-Vasey}, W.~M. and {Xue}, Rui and {Yoachim}, Peter and {Zhang}, Chen and {Zonca}, Andrea and {Astropy Project Contributors}},
        title = "{The Astropy Project: Sustaining and Growing a Community-oriented Open-source Project and the Latest Major Release (v5.0) of the Core Package}",
      journal = {\apj},
         year = 2022,
        month = aug,
       volume = {935},
       number = {2},
          eid = {167},
        pages = {167},
          doi = {10.3847/1538-4357/ac7c74},
archivePrefix = {arXiv},
       eprint = {2206.14220},
 primaryClass = {astro-ph.IM},
       adsurl = {https://ui.adsabs.harvard.edu/abs/2022ApJ...935..167A}
}

@ARTICLE{Aigrain2023,
       author = {{Aigrain}, Suzanne and {Foreman-Mackey}, Daniel},
        title = "{Gaussian Process Regression for Astronomical Time Series}",
      journal = {\araa},
         year = 2023,
        month = aug,
       volume = {61},
        pages = {329-371},
          doi = {10.1146/annurev-astro-052920-103508},
archivePrefix = {arXiv},
       eprint = {2209.08940},
 primaryClass = {astro-ph.IM},
       adsurl = {https://ui.adsabs.harvard.edu/abs/2023ARA&A..61..329A}
}

@ARTICLE{Boisse2011,
       author = {{Boisse}, I. and {Bouchy}, F. and {H{\'e}brard}, G. and {Bonfils}, X. and {Santos}, N. and {Vauclair}, S.},
        title = "{Disentangling between stellar activity and planetary signals}",
      journal = {\aap},
         year = 2011,
        month = apr,
       volume = {528},
          eid = {A4},
        pages = {A4},
          doi = {10.1051/0004-6361/201014354},
       adsurl = {https://ui.adsabs.harvard.edu/abs/2011A&A...528A...4B}
}

@INPROCEEDINGS{atlas9,
       author = {{Castelli}, F. and {Kurucz}, R.~L.},
        title = "{New Grids of ATLAS9 Model Atmospheres}",
    booktitle = {Modelling of Stellar Atmospheres},
         year = 2003,
       editor = {{Piskunov}, N. and {Weiss}, W.~W. and {Gray}, D.~F.},
       volume = {210},
        month = jan,
        pages = {A20},
          doi = {10.48550/arXiv.astro-ph/0405087},
archivePrefix = {arXiv},
       eprint = {astro-ph/0405087},
 primaryClass = {astro-ph},
       adsurl = {https://ui.adsabs.harvard.edu/abs/2003IAUS..210P.A20C}
}

@ARTICLE{carroll12,
       author = {{Carroll}, T.~A. and {Strassmeier}, K.~G. and {Rice}, J.~B. and {K{\"u}nstler}, A.},
        title = "{The magnetic field topology of the weak-lined T Tauri star V410 Tauri. New strategies for Zeeman-Doppler imaging}",
      journal = {\aap},
         year = 2012,
        month = dec,
       volume = {548},
          eid = {A95},
        pages = {A95},
          doi = {10.1051/0004-6361/201220215},
archivePrefix = {arXiv},
       eprint = {1211.2720},
 primaryClass = {astro-ph.SR},
       adsurl = {https://ui.adsabs.harvard.edu/abs/2012A&A...548A..95C}
}

@ARTICLE{dan:bur,
       author = {{Danby}, J.~M.~A. and {Burkardt}, T.~M.},
        title = "{The Solution of Kepler's Equation - Part One}",
      journal = {Celestial Mechanics},
         year = 1983,
        month = oct,
       volume = {31},
       number = {2},
        pages = {95-107},
          doi = {10.1007/BF01686811},
       adsurl = {https://ui.adsabs.harvard.edu/abs/1983CeMec..31...95D}
}

@INPROCEEDINGS{Desort2007,
       author = {{Desort}, M. and {Lagrange}, A. -M. and {Galland}, F. and {Udry}, S. and {Mayor}, M.},
        title = "{Planets and brown dwarfs around A-F type stars: observational results and activity modeling}",
    booktitle = {SF2A-2007: Proceedings of the Annual meeting of the French Society of Astronomy and Astrophysics},
         year = 2007,
       editor = {{Bouvier}, J. and {Chalabaev}, A. and {Charbonnel}, C.},
        month = jul,
        pages = {402},
       adsurl = {https://ui.adsabs.harvard.edu/abs/2007sf2a.conf..402D}
}

@ARTICLE{DR3,
       author = {{Gaia Collaboration} and {Vallenari}, A. and {Brown}, A.~G.~A. and {Prusti}, T. and {de Bruijne}, J.~H.~J. and {Arenou}, F. and {Babusiaux}, C. and {Biermann}, M. and {Creevey}, O.~L. and {Ducourant}, C. and {Evans}, D.~W. and {Eyer}, L. and {Guerra}, R. and {Hutton}, A. and {Jordi}, C. and {Klioner}, S.~A. and {Lammers}, U.~L. and {Lindegren}, L. and {Luri}, X. and {Mignard}, F. and {Panem}, C. and {Pourbaix}, D. and {Randich}, S. and {Sartoretti}, P. and {Soubiran}, C. and {Tanga}, P. and {Walton}, N.~A. and {Bailer-Jones}, C.~A.~L. and {Bastian}, U. and {Drimmel}, R. and {Jansen}, F. and {Katz}, D. and {Lattanzi}, M.~G. and {van Leeuwen}, F. and {Bakker}, J. and {Cacciari}, C. and {Casta{\~n}eda}, J. and {De Angeli}, F. and {Fabricius}, C. and {Fouesneau}, M. and {Fr{\'e}mat}, Y. and {Galluccio}, L. and {Guerrier}, A. and {Heiter}, U. and {Masana}, E. and {Messineo}, R. and {Mowlavi}, N. and {Nicolas}, C. and {Nienartowicz}, K. and {Pailler}, F. and {Panuzzo}, P. and {Riclet}, F. and {Roux}, W. and {Seabroke}, G.~M. and {Sordo}, R. and {Th{\'e}venin}, F. and {Gracia-Abril}, G. and {Portell}, J. and {Teyssier}, D. and {Altmann}, M. and {Andrae}, R. and {Audard}, M. and {Bellas-Velidis}, I. and {Benson}, K. and {Berthier}, J. and {Blomme}, R. and {Burgess}, P.~W. and {Busonero}, D. and {Busso}, G. and {C{\'a}novas}, H. and {Carry}, B. and {Cellino}, A. and {Cheek}, N. and {Clementini}, G. and {Damerdji}, Y. and {Davidson}, M. and {de Teodoro}, P. and {Nu{\~n}ez Campos}, M. and {Delchambre}, L. and {Dell'Oro}, A. and {Esquej}, P. and {Fern{\'a}ndez-Hern{\'a}ndez}, J. and {Fraile}, E. and {Garabato}, D. and {Garc{\'\i}a-Lario}, P. and {Gosset}, E. and {Haigron}, R. and {Halbwachs}, J. -L. and {Hambly}, N.~C. and {Harrison}, D.~L. and {Hern{\'a}ndez}, J. and {Hestroffer}, D. and {Hodgkin}, S.~T. and {Holl}, B. and {Jan{\ss}en}, K. and {Jevardat de Fombelle}, G. and {Jordan}, S. and {Krone-Martins}, A. and {Lanzafame}, A.~C. and {L{\"o}ffler}, W. and {Marchal}, O. and {Marrese}, P.~M. and {Moitinho}, A. and {Muinonen}, K. and {Osborne}, P. and {Pancino}, E. and {Pauwels}, T. and {Recio-Blanco}, A. and {Reyl{\'e}}, C. and {Riello}, M. and {Rimoldini}, L. and {Roegiers}, T. and {Rybizki}, J. and {Sarro}, L.~M. and {Siopis}, C. and {Smith}, M. and {Sozzetti}, A. and {Utrilla}, E. and {van Leeuwen}, M. and {Abbas}, U. and {{\'A}brah{\'a}m}, P. and {Abreu Aramburu}, A. and {Aerts}, C. and {Aguado}, J.~J. and {Ajaj}, M. and {Aldea-Montero}, F. and {Altavilla}, G. and {{\'A}lvarez}, M.~A. and {Alves}, J. and {Anders}, F. and {Anderson}, R.~I. and {Anglada Varela}, E. and {Antoja}, T. and {Baines}, D. and {Baker}, S.~G. and {Balaguer-N{\'u}{\~n}ez}, L. and {Balbinot}, E. and {Balog}, Z. and {Barache}, C. and {Barbato}, D. and {Barros}, M. and {Barstow}, M.~A. and {Bartolom{\'e}}, S. and {Bassilana}, J. -L. and {Bauchet}, N. and {Becciani}, U. and {Bellazzini}, M. and {Berihuete}, A. and {Bernet}, M. and {Bertone}, S. and {Bianchi}, L. and {Binnenfeld}, A. and {Blanco-Cuaresma}, S. and {Blazere}, A. and {Boch}, T. and {Bombrun}, A. and {Bossini}, D. and {Bouquillon}, S. and {Bragaglia}, A. and {Bramante}, L. and {Breedt}, E. and {Bressan}, A. and {Brouillet}, N. and {Brugaletta}, E. and {Bucciarelli}, B. and {Burlacu}, A. and {Butkevich}, A.~G. and {Buzzi}, R. and {Caffau}, E. and {Cancelliere}, R. and {Cantat-Gaudin}, T. and {Carballo}, R. and {Carlucci}, T. and {Carnerero}, M.~I. and {Carrasco}, J.~M. and {Casamiquela}, L. and {Castellani}, M. and {Castro-Ginard}, A. and {Chaoul}, L. and {Charlot}, P. and {Chemin}, L. and {Chiaramida}, V. and {Chiavassa}, A. and {Chornay}, N. and {Comoretto}, G. and {Contursi}, G. and {Cooper}, W.~J. and {Cornez}, T. and {Cowell}, S. and {Crifo}, F. and {Cropper}, M. and {Crosta}, M. and {Crowley}, C. and {Dafonte}, C. and {Dapergolas}, A. and {David}, M. and {David}, P. and {de Laverny}, P. and {De Luise}, F. and {De March}, R. and {De Ridder}, J. and {de Souza}, R. and {de Torres}, A. and {del Peloso}, E.~F. and {del Pozo}, E. and {Delbo}, M. and {Delgado}, A. and {Delisle}, J. -B. and {Demouchy}, C. and {Dharmawardena}, T.~E. and {Di Matteo}, P. and {Diakite}, S. and {Diener}, C. and {Distefano}, E. and {Dolding}, C. and {Edvardsson}, B. and {Enke}, H. and {Fabre}, C. and {Fabrizio}, M. and {Faigler}, S. and {Fedorets}, G. and {Fernique}, P. and {Fienga}, A. and {Figueras}, F. and {Fournier}, Y. and {Fouron}, C. and {Fragkoudi}, F. and {Gai}, M. and {Garcia-Gutierrez}, A. and {Garcia-Reinaldos}, M. and {Garc{\'\i}a-Torres}, M. and {Garofalo}, A. and {Gavel}, A. and {Gavras}, P. and {Gerlach}, E. and {Geyer}, R. and {Giacobbe}, P. and {Gilmore}, G. and {Girona}, S. and {Giuffrida}, G. and {Gomel}, R. and {Gomez}, A. and {Gonz{\'a}lez-N{\'u}{\~n}ez}, J. and {Gonz{\'a}lez-Santamar{\'\i}a}, I. and {Gonz{\'a}lez-Vidal}, J.~J. and {Granvik}, M. and {Guillout}, P. and {Guiraud}, J. and {Guti{\'e}rrez-S{\'a}nchez}, R. and {Guy}, L.~P. and {Hatzidimitriou}, D. and {Hauser}, M. and {Haywood}, M. and {Helmer}, A. and {Helmi}, A. and {Sarmiento}, M.~H. and {Hidalgo}, S.~L. and {Hilger}, T. and {H{\l}adczuk}, N. and {Hobbs}, D. and {Holland}, G. and {Huckle}, H.~E. and {Jardine}, K. and {Jasniewicz}, G. and {Jean-Antoine Piccolo}, A. and {Jim{\'e}nez-Arranz}, {\'O}. and {Jorissen}, A. and {Juaristi Campillo}, J. and {Julbe}, F. and {Karbevska}, L. and {Kervella}, P. and {Khanna}, S. and {Kontizas}, M. and {Kordopatis}, G. and {Korn}, A.~J. and {K{\'o}sp{\'a}l}, {\'A}. and {Kostrzewa-Rutkowska}, Z. and {Kruszy{\'n}ska}, K. and {Kun}, M. and {Laizeau}, P. and {Lambert}, S. and {Lanza}, A.~F. and {Lasne}, Y. and {Le Campion}, J. -F. and {Lebreton}, Y. and {Lebzelter}, T. and {Leccia}, S. and {Leclerc}, N. and {Lecoeur-Taibi}, I. and {Liao}, S. and {Licata}, E.~L. and {Lindstr{\o}m}, H.~E.~P. and {Lister}, T.~A. and {Livanou}, E. and {Lobel}, A. and {Lorca}, A. and {Loup}, C. and {Madrero Pardo}, P. and {Magdaleno Romeo}, A. and {Managau}, S. and {Mann}, R.~G. and {Manteiga}, M. and {Marchant}, J.~M. and {Marconi}, M. and {Marcos}, J. and {Marcos Santos}, M.~M.~S. and {Mar{\'\i}n Pina}, D. and {Marinoni}, S. and {Marocco}, F. and {Marshall}, D.~J. and {Martin Polo}, L. and {Mart{\'\i}n-Fleitas}, J.~M. and {Marton}, G. and {Mary}, N. and {Masip}, A. and {Massari}, D. and {Mastrobuono-Battisti}, A. and {Mazeh}, T. and {McMillan}, P.~J. and {Messina}, S. and {Michalik}, D. and {Millar}, N.~R. and {Mints}, A. and {Molina}, D. and {Molinaro}, R. and {Moln{\'a}r}, L. and {Monari}, G. and {Mongui{\'o}}, M. and {Montegriffo}, P. and {Montero}, A. and {Mor}, R. and {Mora}, A. and {Morbidelli}, R. and {Morel}, T. and {Morris}, D. and {Muraveva}, T. and {Murphy}, C.~P. and {Musella}, I. and {Nagy}, Z. and {Noval}, L. and {Oca{\~n}a}, F. and {Ogden}, A. and {Ordenovic}, C. and {Osinde}, J.~O. and {Pagani}, C. and {Pagano}, I. and {Palaversa}, L. and {Palicio}, P.~A. and {Pallas-Quintela}, L. and {Panahi}, A. and {Payne-Wardenaar}, S. and {Pe{\~n}alosa Esteller}, X. and {Penttil{\"a}}, A. and {Pichon}, B. and {Piersimoni}, A.~M. and {Pineau}, F. -X. and {Plachy}, E. and {Plum}, G. and {Poggio}, E. and {Pr{\v{s}}a}, A. and {Pulone}, L. and {Racero}, E. and {Ragaini}, S. and {Rainer}, M. and {Raiteri}, C.~M. and {Rambaux}, N. and {Ramos}, P. and {Ramos-Lerate}, M. and {Re Fiorentin}, P. and {Regibo}, S. and {Richards}, P.~J. and {Rios Diaz}, C. and {Ripepi}, V. and {Riva}, A. and {Rix}, H. -W. and {Rixon}, G. and {Robichon}, N. and {Robin}, A.~C. and {Robin}, C. and {Roelens}, M. and {Rogues}, H.~R.~O. and {Rohrbasser}, L. and {Romero-G{\'o}mez}, M. and {Rowell}, N. and {Royer}, F. and {Ruz Mieres}, D. and {Rybicki}, K.~A. and {Sadowski}, G. and {S{\'a}ez N{\'u}{\~n}ez}, A. and {Sagrist{\`a} Sell{\'e}s}, A. and {Sahlmann}, J. and {Salguero}, E. and {Samaras}, N. and {Sanchez Gimenez}, V. and {Sanna}, N. and {Santove{\~n}a}, R. and {Sarasso}, M. and {Schultheis}, M. and {Sciacca}, E. and {Segol}, M. and {Segovia}, J.~C. and {S{\'e}gransan}, D. and {Semeux}, D. and {Shahaf}, S. and {Siddiqui}, H.~I. and {Siebert}, A. and {Siltala}, L. and {Silvelo}, A. and {Slezak}, E. and {Slezak}, I. and {Smart}, R.~L. and {Snaith}, O.~N. and {Solano}, E. and {Solitro}, F. and {Souami}, D. and {Souchay}, J. and {Spagna}, A. and {Spina}, L. and {Spoto}, F. and {Steele}, I.~A. and {Steidelm{\"u}ller}, H. and {Stephenson}, C.~A. and {S{\"u}veges}, M. and {Surdej}, J. and {Szabados}, L. and {Szegedi-Elek}, E. and {Taris}, F. and {Taylor}, M.~B. and {Teixeira}, R. and {Tolomei}, L. and {Tonello}, N. and {Torra}, F. and {Torra}, J. and {Torralba Elipe}, G. and {Trabucchi}, M. and {Tsounis}, A.~T. and {Turon}, C. and {Ulla}, A. and {Unger}, N. and {Vaillant}, M.~V. and {van Dillen}, E. and {van Reeven}, W. and {Vanel}, O. and {Vecchiato}, A. and {Viala}, Y. and {Vicente}, D. and {Voutsinas}, S. and {Weiler}, M. and {Wevers}, T. and {Wyrzykowski}, {\L}. and {Yoldas}, A. and {Yvard}, P. and {Zhao}, H. and {Zorec}, J. and {Zucker}, S. and {Zwitter}, T.},
        title = "{Gaia Data Release 3. Summary of the content and survey properties}",
      journal = {\aap},
         year = 2023,
        month = jun,
       volume = {674},
          eid = {A1},
        pages = {A1},
          doi = {10.1051/0004-6361/202243940},
archivePrefix = {arXiv},
       eprint = {2208.00211},
 primaryClass = {astro-ph.GA},
       adsurl = {https://ui.adsabs.harvard.edu/abs/2023A&A...674A...1G}
}

@ARTICLE{Donati1997,
       author = {{Donati}, J.-F. and {Semel}, M. and {Carter}, B.~D. and {Rees}, D.~E. and {Collier Cameron}, A.},
        title = "{Spectropolarimetric observations of active stars}",
      journal = {\mnras},
         year = 1997,
        month = nov,
       volume = {291},
       number = {4},
        pages = {658-682},
          doi = {10.1093/mnras/291.4.658},
       adsurl = {https://ui.adsabs.harvard.edu/abs/1997MNRAS.291..658D}
}

@dataset{Ducati2002,
       author = {{Ducati}, J.~R.},
        title = "{VizieR Online Data Catalog: Catalogue of Stellar Photometry in Johnson's 11-color system.}",
 howpublished = {CDS/ADC Collection of Electronic Catalogues, 2237, 0 (2002)},
         year = 2002,
        month = jan,
       adsurl = {https://ui.adsabs.harvard.edu/abs/2002yCat.2237....0D}
}

@ARTICLE{Eker1989,
       author = {{Eker}, Zeki},
        title = "{High-resolution H{\ensuremath{\alpha}} spectroscopy of the bright RS CVn system HR 7275.}",
      journal = {\mnras},
         year = 1989,
        month = may,
       volume = {238},
        pages = {675-688},
          doi = {10.1093/mnras/238.2.675},
       adsurl = {https://ui.adsabs.harvard.edu/abs/1989MNRAS.238..675E}
}

@ARTICLE{Fried1982,
       author = {{Fried}, R.~E. and {Eaton}, Joel A. and {Hall}, D.~S. and {Henry}, G.~W. and {Lovell}, L.~P. and {Krisciunas}, K. and {Chambliss}, C.~R. and {Detterline}, P.~K. and {Landis}, H.~J. and {Louth}, H. and {Skillman}, D.~R.},
        title = "{HR7275 - a New Variable Star}",
      journal = {\apss},
         year = 1982,
        month = apr,
       volume = {83},
       number = {1-2},
        pages = {181-188},
          doi = {10.1007/BF00648550},
       adsurl = {https://ui.adsabs.harvard.edu/abs/1982Ap&SS..83..181F}
}

@ARTICLE{Folsom2010,
       author = {{Folsom}, C.~P. and {Kochukhov}, O. and {Wade}, G.~A. and {Silvester}, J. and {Bagnulo}, S.},
        title = "{Magnetic field, chemical composition and line profile variability of the peculiar eclipsing binary star AR Aur}",
      journal = {\mnras},
         year = 2010,
        month = oct,
       volume = {407},
       number = {4},
        pages = {2383-2392},
          doi = {10.1111/j.1365-2966.2010.17057.x},
archivePrefix = {arXiv},
       eprint = {1005.3793},
 primaryClass = {astro-ph.SR},
       adsurl = {https://ui.adsabs.harvard.edu/abs/2010MNRAS.407.2383F}
}

@BOOK{Gray2022,
       author = {{Gray}, David F.},
        title = "{The observation and analysis of stellar photospheres}",
         year = 2022,
          doi = {10.1017/9781009082136},
       adsurl = {https://ui.adsabs.harvard.edu/abs/2022oasp.book.....G}
}

@ARTICLE{Hall1972,
       author = {{Hall}, Douglas S.},
        title = "{A T Tauri-Like Star in the Eclipsing Binary RS Canum Venaticorum}",
      journal = {\pasp},
         year = 1972,
        month = apr,
       volume = {84},
       number = {498},
        pages = {323},
          doi = {10.1086/129291},
       adsurl = {https://ui.adsabs.harvard.edu/abs/1972PASP...84..323H}
}

@ARTICLE{Hatzes2002,
       author = {{Hatzes}, A.~P.},
        title = "{Starspots and exoplanets}",
      journal = {Astronomische Nachrichten},
         year = 2002,
        month = jul,
       volume = {323},
        pages = {392-394},
          doi = {10.1002/1521-3994(200208)323:3/4<392::AID-ASNA392>3.0.CO;2-M},
       adsurl = {https://ui.adsabs.harvard.edu/abs/2002AN....323..392H}
}

@ARTICLE{numpyharris2020,
 title         = {Array programming with {NumPy}},
 author        = {Charles R. Harris and K. Jarrod Millman and St{\'{e}}fan J.
                 van der Walt and Ralf Gommers and Pauli Virtanen and David
                 Cournapeau and Eric Wieser and Julian Taylor and Sebastian
                 Berg and Nathaniel J. Smith and Robert Kern and Matti Picus
                 and Stephan Hoyer and Marten H. van Kerkwijk and Matthew
                 Brett and Allan Haldane and Jaime Fern{\'{a}}ndez del
                 R{\'{i}}o and Mark Wiebe and Pearu Peterson and Pierre
                 G{\'{e}}rard-Marchant and Kevin Sheppard and Tyler Reddy and
                 Warren Weckesser and Hameer Abbasi and Christoph Gohlke and
                 Travis E. Oliphant},
 year          = {2020},
 month         = sep,
 journal       = {Nature},
 volume        = {585},
 number        = {7825},
 pages         = {357--362},
 doi           = {10.1038/s41586-020-2649-2},
 publisher     = {Springer Science and Business Media {LLC}},
 url           = {https://doi.org/10.1038/s41586-020-2649-2}
}

@ARTICLE{Hensberge2008,
       author = {{Hensberge}, H. and {Iliji{\'c}}, S. and {Torres}, K.~B.~V.},
        title = "{On the separation of component spectra in binary and higher-multiplicity stellar systems: bias progression and spurious patterns}",
      journal = {\aap},
         year = 2008,
        month = may,
       volume = {482},
       number = {3},
        pages = {1031-1051},
          doi = {10.1051/0004-6361:20079038},
       adsurl = {https://ui.adsabs.harvard.edu/abs/2008A&A...482.1031H}
}

@ARTICLE{Ilyin2000,
       author = {{Ilyin}, I.~V.},
        title = "{High resolution SOFIN CCD echelle spectroscopy}",
      journal = {University of Oulu, Division of Astronomy},
          year = 2000,
        month = sep,
       adsurl = {https://ui.adsabs.harvard.edu/abs/2000PhDT..........I}
}

@ARTICLE{ilyin2012,
       author = {{Ilyin}, I.},
        title = "{Second-order error propagation in the Mueller matrix of a  spectropolarimeter}",
      journal = {Astronomische Nachrichten},
         year = 2012,
        month = apr,
       volume = {333},
       number = {3},
        pages = {213},
          doi = {10.1002/asna.201211652},
       adsurl = {https://ui.adsabs.harvard.edu/abs/2012AN....333..213I}
}

@ARTICLE{Jarvinen2025,
       author = {{J{\"a}rvinen}, S.~P. and {Strassmeier}, K.~G.},
        title = "{A search for Maunder-minimum candidate stars}",
      journal = {\aap},
         year = 2025,
        month = jun,
       volume = {698},
          eid = {A93},
        pages = {A93},
          doi = {10.1051/0004-6361/202554111},
archivePrefix = {arXiv},
       eprint = {2504.19670},
 primaryClass = {astro-ph.SR},
       adsurl = {https://ui.adsabs.harvard.edu/abs/2025A&A...698A..93J}
}

@ARTICLE{Kochukhov2010,
       author = {{Kochukhov}, O. and {Makaganiuk}, V. and {Piskunov}, N.},
        title = "{Least-squares deconvolution of the stellar intensity and polarization spectra}",
      journal = {\aap},
         year = 2010,
        month = dec,
       volume = {524},
          eid = {A5},
        pages = {A5},
          doi = {10.1051/0004-6361/201015429},
archivePrefix = {arXiv},
       eprint = {1008.5115},
 primaryClass = {astro-ph.SR},
       adsurl = {https://ui.adsabs.harvard.edu/abs/2010A&A...524A...5K}
}

@ARTICLE{Kovari2013,
       author = {{K{\H{o}}v{\'a}ri}, {\mbox Zs}. and {Korhonen}, H. and {Strassmeier}, K.~G. and {Weber}, M. and {Kriskovics}, L. and {Savanov}, I.},
        title = "{Doppler imaging of stellar surface structure. XXIV. The lithium-rich single K-giants DP Canum Venaticorum and DI Piscium}",
      journal = {\aap},
         year = 2013,
        month = mar,
       volume = {551},
          eid = {A2},
        pages = {A2},
          doi = {10.1051/0004-6361/201220227},
archivePrefix = {arXiv},
       eprint = {1301.0445},
 primaryClass = {astro-ph.SR},
       adsurl = {https://ui.adsabs.harvard.edu/abs/2013A&A...551A...2K}
}

@ARTICLE{Kovari2024,
       author = {{K{\H{o}}v{\'a}ri}, {\mbox Zs}. and {Strassmeier}, K.~G. and {Kriskovics}, L. and {Ol{\'a}h}, K. and {Borkovits}, T. and {Radv{\'a}nyi}, {\'A}. and {Granzer}, T. and {Seli}, B. and {Vida}, K. and {Weber}, M.},
        title = "{A star under multiple influences. Magnetic activity in V815 Her, a compact 2+2 hierarchical system}",
      journal = {\aap},
         year = 2024,
        month = apr,
       volume = {684},
          eid = {A94},
        pages = {A94},
          doi = {10.1051/0004-6361/202348324},
archivePrefix = {arXiv},
       eprint = {2312.08416},
 primaryClass = {astro-ph.SR},
       adsurl = {https://ui.adsabs.harvard.edu/abs/2024A&A...684A..94K}
}

@ARTICLE{Kriskovics2013,
       author = {{Kriskovics}, L. and {Vida}, K. and {K{\H{o}}v{\'a}ri}, {\mbox Zs}. and {Garcia-Alvarez}, D. and {Ol{\'a}h}, K.},
        title = "{Doppler imaging of the double-lined active binary V824 Ara}",
      journal = {Astronomische Nachrichten},
         year = 2013,
        month = nov,
       volume = {334},
       number = {9},
        pages = {976},
          doi = {10.1002/asna.201211974},
archivePrefix = {arXiv},
       eprint = {1310.4020},
 primaryClass = {astro-ph.SR},
       adsurl = {https://ui.adsabs.harvard.edu/abs/2013AN....334..976K}
}

@ARTICLE{Kurucz1993,
       author = {{Kurucz}, Robert},
        title = "{ATLAS9 Stellar Atmosphere Programs and 2 km/s grid.}",
      journal = {Robert Kurucz CD-ROM},
         year = 1993,
        month = jan,
       volume = {13},
       adsurl = {https://ui.adsabs.harvard.edu/abs/1993KurCD..13.....K}
}

@ARTICLE{Leconte2011,
       author = {{Leconte}, J. and {Lai}, D. and {Chabrier}, G.},
        title = "{Distorted, nonspherical transiting planets: impact on the transit depth and on the radius determination}",
      journal = {\aap},
         year = 2011,
        month = apr,
       volume = {528},
          eid = {A41},
        pages = {A41},
          doi = {10.1051/0004-6361/201015811},
archivePrefix = {arXiv},
       eprint = {1101.2813},
 primaryClass = {astro-ph.EP},
       adsurl = {https://ui.adsabs.harvard.edu/abs/2011A&A...528A..41L}
}

@ARTICLE{Marcs2008,
       author = {{Gustafsson}, B. and {Edvardsson}, B. and {Eriksson}, K. and {J{\o}rgensen}, U.~G. and {Nordlund}, {\r{A}}. and {Plez}, B.},
        title = "{A grid of MARCS model atmospheres for late-type stars. I. Methods and general properties}",
      journal = {\aap},
         year = 2008,
        month = aug,
       volume = {486},
       number = {3},
        pages = {951-970},
          doi = {10.1051/0004-6361:200809724},
archivePrefix = {arXiv},
       eprint = {0805.0554},
 primaryClass = {astro-ph},
       adsurl = {https://ui.adsabs.harvard.edu/abs/2008A&A...486..951G}
}

@ARTICLE{Medeiros1999,
       author = {{de~Medeiros}, J.~R. and {Udry}, S.},
        title = "{Ten CORAVEL spectroscopic binary orbits of evolved stars}",
      journal = {\aap},
         year = 1999,
        month = jun,
       volume = {346},
        pages = {532-536},
       adsurl = {https://ui.adsabs.harvard.edu/abs/1999A&A...346..532D}
}

@INPROCEEDINGS{Menuier2023,
       author = {{Menuier}, Nad{\`e}ge},
        title = "{Stellar variability in radial velocity}",
      journal = {\edpsp},
    booktitle = {Star-Planet Interactions},
         year = 2023,
       editor = {{Bigot}, Lionel and {Bouvier}, J{\'e}r{\^o}me and {Lebreton}, Yveline and {Chiavassa}, Andrea and {L{\`e}bre}, Agn{\`e}s},
        month = feb,
        pages = {22},
          doi = {10.48550/arXiv.2104.06072},
archivePrefix = {arXiv},
       eprint = {2104.06072},
 primaryClass = {astro-ph.SR},
       adsurl = {https://ui.adsabs.harvard.edu/abs/2023spi..conf...22M}
}

@ARTICLE{Melendez2012,
       author = {{Mel{\'e}ndez}, J. and {Bergemann}, M. and {Cohen}, J.~G. and {Endl}, M. and {Karakas}, A.~I. and {Ram{\'\i}rez}, I. and {Cochran}, W.~D. and {Yong}, D. and {MacQueen}, P.~J. and {Kobayashi}, C. and {Asplund}, M.},
        title = "{The remarkable solar twin HIP 56948: a prime target in the quest for other Earths}",
      journal = {\aap},
         year = 2012,
        month = jul,
       volume = {543},
          eid = {A29},
        pages = {A29},
          doi = {10.1051/0004-6361/201117222},
archivePrefix = {arXiv},
       eprint = {1204.2766},
 primaryClass = {astro-ph.SR},
       adsurl = {https://ui.adsabs.harvard.edu/abs/2012A&A...543A..29M}
}

@ARTICLE{Neff1995,
       author = {{Neff}, James E. and {O'Neal}, Douglas and {Saar}, Steven H.},
        title = "{Absolute Measurements of Starspot Area and Temperature: II Pegasi in 1989 October}",
      journal = {\apj},
         year = 1995,
        month = oct,
       volume = {452},
        pages = {879},
          doi = {10.1086/176356},
       adsurl = {https://ui.adsabs.harvard.edu/abs/1995ApJ...452..879N}
}

@ARTICLE{Osten1998,
       author = {{Osten}, R.~A. and {Saar}, S.~H.},
        title = "{Physical properties of active stars and stellar systems}",
      journal = {\mnras},
         year = 1998,
        month = apr,
       volume = {295},
       number = {2},
        pages = {257-264},
          doi = {10.1046/j.1365-8711.1998.01121.x},
       adsurl = {https://ui.adsabs.harvard.edu/abs/1998MNRAS.295..257O}
}

@ARTICLE{parses,
       author = {{Jovanovic}, M. and {Weber}, M. and {Allende Prieto}, C.},
        title = "{Parses Pipeline For Determining The Stellar Parameters}",
      journal = {Publications de l'Observatoire Astronomique de Beograd},
         year = 2013,
        month = may,
       volume = {92},
        pages = {169-174},
       adsurl = {https://ui.adsabs.harvard.edu/abs/2013POBeo..92..169J}
}

@ARTICLE{pis:weh,
       author = {{Piskunov}, N.~E. and {Wehlau}, W.~H.},
        title = "{Mapping stellar surfaces from spectra of medium resolution}",
      journal = {\aap},
         year = 1990,
        month = jul,
       volume = {233},
       number = {2},
        pages = {497-502},
       adsurl = {https://ui.adsabs.harvard.edu/abs/1990A&A...233..497P}
}

@ARTICLE{vald3,
       author = {{Ryabchikova}, T. and {Piskunov}, N. and {Kurucz}, R.~L. and {Stempels}, H.~C. and {Heiter}, U. and {Pakhomov}, Yu and {Barklem}, P.~S.},
        title = "{A major upgrade of the VALD database}",
      journal = {\physscr},
         year = 2015,
        month = may,
       volume = {90},
       number = {5},
          eid = {054005},
        pages = {054005},
          doi = {10.1088/0031-8949/90/5/054005},
       adsurl = {https://ui.adsabs.harvard.edu/abs/2015PhyS...90e4005R}
}

@ARTICLE{Popper1980,
       author = {{Popper}, D.~M.},
        title = "{Stellar masses.}",
      journal = {\araa},
         year = 1980,
        month = jan,
       volume = {18},
        pages = {115-164},
          doi = {10.1146/annurev.aa.18.090180.000555},
       adsurl = {https://ui.adsabs.harvard.edu/abs/1980ARA&A..18..115P}
}

@ARTICLE{Saar&Donahue1997,
       author = {{Saar}, Steven H. and {Donahue}, Robert A.},
        title = "{Activity-Related Radial Velocity Variation in Cool Stars}",
      journal = {\apj},
         year = 1997,
        month = aug,
       volume = {485},
       number = {1},
        pages = {319-327},
          doi = {10.1086/304392},
       adsurl = {https://ui.adsabs.harvard.edu/abs/1997ApJ...485..319S}
}

@ARTICLE{Sablowski2019,
       author = {{Sablowski}, Daniel P. and {J{\"a}rvinen}, Silva and {Weber}, Michael},
        title = "{Spectangular: Disentangling variable spectra}",
      journal = {\aap},
         year = 2019,
        month = mar,
       volume = {623},
          eid = {A31},
        pages = {A31},
          doi = {10.1051/0004-6361/201834836},
archivePrefix = {arXiv},
       eprint = {1902.00318},
 primaryClass = {astro-ph.IM},
       adsurl = {https://ui.adsabs.harvard.edu/abs/2019A&A...623A..31S}
}

@ARTICLE{SciPy2020-NMeth,
  author  = {Virtanen, Pauli and Gommers, Ralf and Oliphant, Travis E. and
            Haberland, Matt and Reddy, Tyler and Cournapeau, David and
            Burovski, Evgeni and Peterson, Pearu and Weckesser, Warren and
            Bright, Jonathan and {van der Walt}, St{\'e}fan J. and
            Brett, Matthew and Wilson, Joshua and Millman, K. Jarrod and
            Mayorov, Nikolay and Nelson, Andrew R. J. and Jones, Eric and
            Kern, Robert and Larson, Eric and Carey, C J and
            Polat, {\.I}lhan and Feng, Yu and Moore, Eric W. and
            {VanderPlas}, Jake and Laxalde, Denis and Perktold, Josef and
            Cimrman, Robert and Henriksen, Ian and Quintero, E. A. and
            Harris, Charles R. and Archibald, Anne M. and
            Ribeiro, Ant{\^o}nio H. and Pedregosa, Fabian and
            {van Mulbregt}, Paul and {SciPy 1.0 Contributors}},
  title   = {{{SciPy} 1.0: Fundamental Algorithms for Scientific
            Computing in Python}},
  journal = {Nature Methods},
  year    = {2020},
  volume  = {17},
  pages   = {261--272},
  adsurl  = {https://rdcu.be/b08Wh},
  doi     = {10.1038/s41592-019-0686-2},
}

@ARTICLE{Sol&Ste1984,
       author = {{Solanki}, S.~K. and {Stenflo}, J.~O.},
        title = "{Properties of solar magnetic fluxtubes as revealed by Fe I lines}",
      journal = {\aap},
         year = 1984,
        month = nov,
       volume = {140},
       number = {1},
        pages = {185-198},
       adsurl = {https://ui.adsabs.harvard.edu/abs/1984A&A...140..185S}
}

@ARTICLE{Steffen2015,
       author = {{Steffen}, M. and {Prakapavi{\v{c}}ius}, D. and {Caffau}, E. and {Ludwig}, H. -G. and {Bonifacio}, P. and {Cayrel}, R. and {Ku{\v{c}}inskas}, A. and {Livingston}, W.~C.},
        title = "{The photospheric solar oxygen project. IV. 3D-NLTE investigation of the 777 nm triplet lines}",
      journal = {\aap},
         year = 2015,
        month = nov,
       volume = {583},
          eid = {A57},
        pages = {A57},
          doi = {10.1051/0004-6361/201526406},
archivePrefix = {arXiv},
       eprint = {1508.03487},
 primaryClass = {astro-ph.SR},
       adsurl = {https://ui.adsabs.harvard.edu/abs/2015A&A...583A..57S}
}

@ARTICLE{stella2004,
       author = {{Strassmeier}, K.~G. and {Granzer}, T. and {Weber}, M. and {Woche}, M. and {Andersen}, M.~I. and {Bartus}, J. and {Bauer}, S. -M. and {Dionies}, F. and {Popow}, E. and {Fechner}, T. and {Hildebrandt}, G. and {Washuettl}, A. and {Ritter}, A. and {Schwope}, A. and {Staude}, A. and {Paschke}, J. and {Stolz}, P.~A. and {Serre-Ricart}, M. and {de la Rosa}, T. and {Arnay}, R.},
        title = "{The STELLA robotic observatory}",
      journal = {Astronomische Nachrichten},
         year = 2004,
        month = oct,
       volume = {325},
       number = {6},
        pages = {527-532},
          doi = {10.1002/asna.200410273},
       adsurl = {https://ui.adsabs.harvard.edu/abs/2004AN....325..527S}
}

@ARTICLE{Strassmeier1989,
       author = {{Strassmeier}, Klaus G. and {Hall}, Douglas S. and {Boyd}, Louis J. and {Genet}, Russell M.},
        title = "{Photometric Variability in Chromospherically Active Stars. III. The Binary Stars}",
      journal = {\apjs},
         year = 1989,
        month = jan,
       volume = {69},
        pages = {141},
          doi = {10.1086/191310},
       adsurl = {https://ui.adsabs.harvard.edu/abs/1989ApJS...69..141S}
}

@ARTICLE{Strassmeier1993,
       author = {{Strassmeier}, K.~G. and {Hall}, D.~S. and {Fekel}, F.~C. and {Scheck}, M.},
        title = "{A catalog of chromospherically active binary stars (second edition).}",
      journal = {\aaps},
         year = 1993,
        month = jul,
       volume = {100},
        pages = {173-225},
       adsurl = {https://ui.adsabs.harvard.edu/abs/1993A&AS..100..173S}
}

@ARTICLE{Strassmeier2009,
       author = {{Strassmeier}, Klaus G.},
        title = "{Starspots}",
      journal = {\aapr},
         year = 2009,
        month = sep,
       volume = {17},
       number = {3},
        pages = {251-308},
          doi = {10.1007/s00159-009-0020-6},
       adsurl = {https://ui.adsabs.harvard.edu/abs/2009A&ARv..17..251S}
}

@ARTICLE{Strassmeier2015,
       author = {{Strassmeier}, K.~G. and {Ilyin}, I. and {J{\"a}rvinen}, A. and {Weber}, M. and {Woche}, M. and {Barnes}, S.~I. and {Bauer}, S. -M. and {Beckert}, E. and {Bittner}, W. and {Bredthauer}, R. and {Carroll}, T.~A. and {Denker}, C. and {Dionies}, F. and {DiVarano}, I. and {D{\"o}scher}, D. and {Fechner}, T. and {Feuerstein}, D. and {Granzer}, T. and {Hahn}, T. and {Harnisch}, G. and {Hofmann}, A. and {Lesser}, M. and {Paschke}, J. and {Pankratow}, S. and {Plank}, V. and {Pl{\"u}schke}, D. and {Popow}, E. and {Sablowski}, D.},
        title = "{PEPSI: The high-resolution {\'e}chelle spectrograph and polarimeter for the Large Binocular Telescope}",
      journal = {Astronomische Nachrichten},
         year = 2015,
        month = may,
       volume = {336},
       number = {4},
        pages = {324},
          doi = {10.1002/asna.201512172},
archivePrefix = {arXiv},
       eprint = {1505.06492},
 primaryClass = {astro-ph.IM},
       adsurl = {https://ui.adsabs.harvard.edu/abs/2015AN....336..324S}
}

@ARTICLE{Strassmeier2018,
       author = {{Strassmeier}, K.~G. and {Ilyin}, I. and {Steffen}, M.},
        title = "{PEPSI deep spectra. I. The Sun-as-a-star}",
      journal = {\aap},
         year = 2018,
        month = apr,
       volume = {612},
          eid = {A44},
        pages = {A44},
          doi = {10.1051/0004-6361/201731631},
archivePrefix = {arXiv},
       eprint = {1712.06960},
 primaryClass = {astro-ph.SR},
       adsurl = {https://ui.adsabs.harvard.edu/abs/2018A&A...612A..44S}
}

@ARTICLE{Strassmeier2022,
       author = {{Strassmeier}, Klaus G. and {Steffen}, Matthias},
        title = "{On the lithium abundance of the visual binary components {\ensuremath{\xi}} Boo A ( G8V ) and {\ensuremath{\xi}} Boo B ( K5V )}",
      journal = {Astronomische Nachrichten},
         year = 2022,
        month = nov,
       volume = {343},
       number = {9-10},
          eid = {e20220036},
        pages = {e20220036},
          doi = {10.1002/asna.20220036},
archivePrefix = {arXiv},
       eprint = {2211.06336},
 primaryClass = {astro-ph.SR},
       adsurl = {https://ui.adsabs.harvard.edu/abs/2022AN....34320036S}
}

@ARTICLE{Strassmeier2024,
       author = {{Strassmeier}, K.~G. and {K{\H{o}}v{\'a}ri}, {\mbox Zs}. and {Weber}, M. and {Granzer}, T.},
        title = "{Long-term Doppler imaging of the star XX Trianguli indicates chaotic non-periodic dynamo}",
      journal = {Nature Communications},
         year = 2024,
        month = dec,
       volume = {15},
       number = {1},
          eid = {9986},
        pages = {9986},
          doi = {10.1038/s41467-024-54329-4},
       adsurl = {https://ui.adsabs.harvard.edu/abs/2024NatCo..15.9986S}
}

@ARTICLE{vpnep,
       author = {{Strassmeier}, K.~G. and {Weber}, M. and {Gruner}, D. and {Ilyin}, I. and {Steffen}, M. and {Baratella}, M. and {J{\"a}rvinen}, S. and {Granzer}, T. and {Barnes}, S.~A. and {Carroll}, T.~A. and {Mallonn}, M. and {Sablowski}, D. and {Gabor}, P. and {Brown}, D. and {Corbally}, C. and {Franz}, M.},
        title = "{VPNEP: Detailed characterization of TESS targets around the Northern Ecliptic Pole. I. Survey design, pilot analysis, and initial data release}",
      journal = {\aap},
         year = 2023,
        month = mar,
       volume = {671},
          eid = {A7},
        pages = {A7},
          doi = {10.1051/0004-6361/202245255},
archivePrefix = {arXiv},
       eprint = {2302.01794},
 primaryClass = {astro-ph.SR},
       adsurl = {https://ui.adsabs.harvard.edu/abs/2023A&A...671A...7S}
}

@software{turbo,
       author = {{Plez}, B.},
        title = "{Turbospectrum: Code for spectral synthesis}",
 howpublished = {Astrophysics Source Code Library, record ascl:1205.004},
         year = 2012,
        month = may,
          eid = {ascl:1205.004},
       adsurl = {https://ui.adsabs.harvard.edu/abs/2012ascl.soft05004P}
}

@ARTICLE{Vogt1983,
       author = {{Vogt}, S.~S. and {Penrod}, G.~D.},
        title = "{Doppler imaging of spotted stars : application to the RS Canum Venaticorum star HR 1099.}",
      journal = {\pasp},
         year = 1983,
        month = sep,
       volume = {95},
        pages = {565-576},
          doi = {10.1086/131208},
       adsurl = {https://ui.adsabs.harvard.edu/abs/1983PASP...95..565V}
}

@ARTICLE{Weber2011,
       author = {{Weber}, M. and {Strassmeier}, K.~G.},
        title = "{The spectroscopic orbit of Capella revisited}",
      journal = {\aap},
         year = 2011,
        month = jul,
       volume = {531},
          eid = {A89},
        pages = {A89},
          doi = {10.1051/0004-6361/201116885},
archivePrefix = {arXiv},
       eprint = {1104.0342},
 primaryClass = {astro-ph.SR},
       adsurl = {https://ui.adsabs.harvard.edu/abs/2011A&A...531A..89W}
}

@INPROCEEDINGS{stella10,
       author = {{Weber}, Michael and {Granzer}, Thomas and {Strassmeier}, Klaus G.},
        title = "{STELLA: 10 years of robotic observations on Tenerife}",
    booktitle = {Observatory Operations: Strategies, Processes, and Systems VI},
         year = 2016,
       editor = {{Peck}, Alison B. and {Seaman}, Robert L. and {Benn}, Chris R.},
       series = {Society of Photo-Optical Instrumentation Engineers (SPIE) Conference Series},
       volume = {9910},
        month = jul,
          eid = {99100N},
        pages = {99100N},
          doi = {10.1117/12.2232251},
       adsurl = {https://ui.adsabs.harvard.edu/abs/2016SPIE.9910E..0NW}
}

@ARTICLE{Young1944,
       author = {{Young}, R.~K.},
        title = "{The Orbit of the Spectroscopic Binary H.D. 179094}",
      journal = {\jrasc},
         year = 1944,
        month = nov,
       volume = {38},
        pages = {366},
       adsurl = {https://ui.adsabs.harvard.edu/abs/1944JRASC..38..366Y}
}

@ARTICLE{Zhao2023,
       author = {{Zhao}, Y. and {Dumusque}, X.},
        title = "{SOAP-GPU: Efficient spectral modeling of stellar activity using graphical processing units}",
      journal = {\aap},
         year = 2023,
        month = mar,
       volume = {671},
          eid = {A11},
        pages = {A11},
          doi = {10.1051/0004-6361/202244568},
archivePrefix = {arXiv},
       eprint = {2301.04259},
 primaryClass = {astro-ph.SR},
       adsurl = {https://ui.adsabs.harvard.edu/abs/2023A&A...671A..11Z}
}

\begin{appendix}

\onecolumn

\section{Extra tables}

\begin{table*}[h!]
\caption{Observing log for STELLA-SES data.}
\begin{tabular}{ccccccc}
\hline\hline\noalign{\smallskip}
Date (UT) & BJD &Orbital Phase & $V_{\rm A}$ [\kms] & $V_{\rm B}$ [\kms] & $O-C$ [\ms] & S/N  \\
\noalign{\smallskip}\hline\noalign{\smallskip}
2022-03-01   &2459640.705    &0.581   &45.895	   &$-$39.906	 &337.101    & 257  \\
2022-03-04   &2459643.713    &0.686   &35.085	   &$-$28.993	 &$-$92.433  & 247   \\
2022-03-06   &2459645.697    &0.756   &20.870	   &$-$12.878	 &$-$44.386  & 311    \\

\dots \\                                                               
\dots \\

2022-11-09   &2459893.322	&0.419    &$-$28.937	   &35.485	&$-$0.0381  & 327   \\
2022-11-10   &2459894.320	&0.454    &$-$37.279	   &42.412	&$-$0.647   & 354    \\
2022-11-15   &2459900.320	&0.664    &$-$33.167	   &38.666	&$-$0.108   & 347     \\
\noalign{\smallskip}\hline
\end{tabular}
\tablefoot{Complete table is available at the CDS.}
\label{Table_STELLA-SES}
\end{table*}

\begin{table*}[h!]
\caption{Observing log for PEPSI-VATT data.}
\begin{tabular}{ccccccc}
\hline\hline\noalign{\smallskip}
Date (UT) & BJD &Orbital Phase & $V_{\rm A}$ [\kms] & $V_{\rm B}$ [\kms] & $O-C$ [\ms]  & S/N  \\
\noalign{\smallskip}\hline\noalign{\smallskip}
2022-05-10   &2459709.827    &0.002   &$-$30.964    &45.862   &215.251     & 408  \\
2022-05-10   &2459709.891    &0.035   &$-$31.217    &45.608   &201.551     & 415 \\
2022-05-11   &2459710.843    &0.037   &$-$34.147    &48.685   &$-$17.853   & 406 \\

\dots \\                                                               
\dots \\

2022-06-07   &2459737.903  	&0.982    &$-$28.824	&43.197	&225.099       & 405    \\
2022-06-13   &2459743.967	&0.194    &$-$23.766	&36.844	&$-$399.837    & 422     \\
2022-06-14   &2459744.836	&0.224    &$-$17.650	&30.947	&$-$259.443    & 716     \\
\noalign{\smallskip}\hline
\end{tabular}
\tablefoot{Complete table is available at the CDS.}
\label{Table_PEPSI-VATT}
\end{table*}

\begin{table}[h!]
\caption{Period analysis of HR\,7275 for different activity indicators.}\label{Period_Analysis}
\begin{tabular}{lrr}
\hline\hline\noalign{\smallskip}
Activity              &$P_{\rm LS}$          &$P_{\rm PDM}$ \\
Indicator             &(days)            &(days)   \\
\noalign{\smallskip}\hline\noalign{\smallskip}
Ca\,{\sc ii}\,IRT-2    & 28.0 $\pm$ 0.1    &  28.0 $\pm$ 0.2   \\
Ca\,{\sc ii}\,H        & 27.8 $\pm$ 0.1    &  27.8 $\pm$ 0.3   \\
Ca\,{\sc ii}\,K       & 27.8 $\pm$ 0.1    &  27.8 $\pm$ 0.3   \\
\Halpha               & 14.2 $\pm$ 0.4    &  14.2 $\pm$ 0.5   \\
$\Delta T$             & 28.2 $\pm$ 0.1    & 14.2 $\pm$ 0.1   \\
RV residuals          & 13.6 $\pm$ 0.3    &  27.2 $\pm$ 0.4   \\
\noalign{\smallskip}\hline
\end{tabular}
\tablefoot{Rotational period is determined by using different activity indicators with two methods; Lomb-Scargle (LS) and phase dispersion minimization (PDM). While the Ca\,{\sc ii}\,IRT and H\&K lines demonstrate emission modulation with approximately the orbital period ($\simeq$28 days), the \Halpha\ and RV residuals show half-period ($\simeq$ 14 days) due to the configuration by multiple spots.}
\end{table}

\section{Extra figures}\label{extrafigs}
\onecolumn 
\begin{figure*}[h!]
    \centering
    \includegraphics[width=16cm]{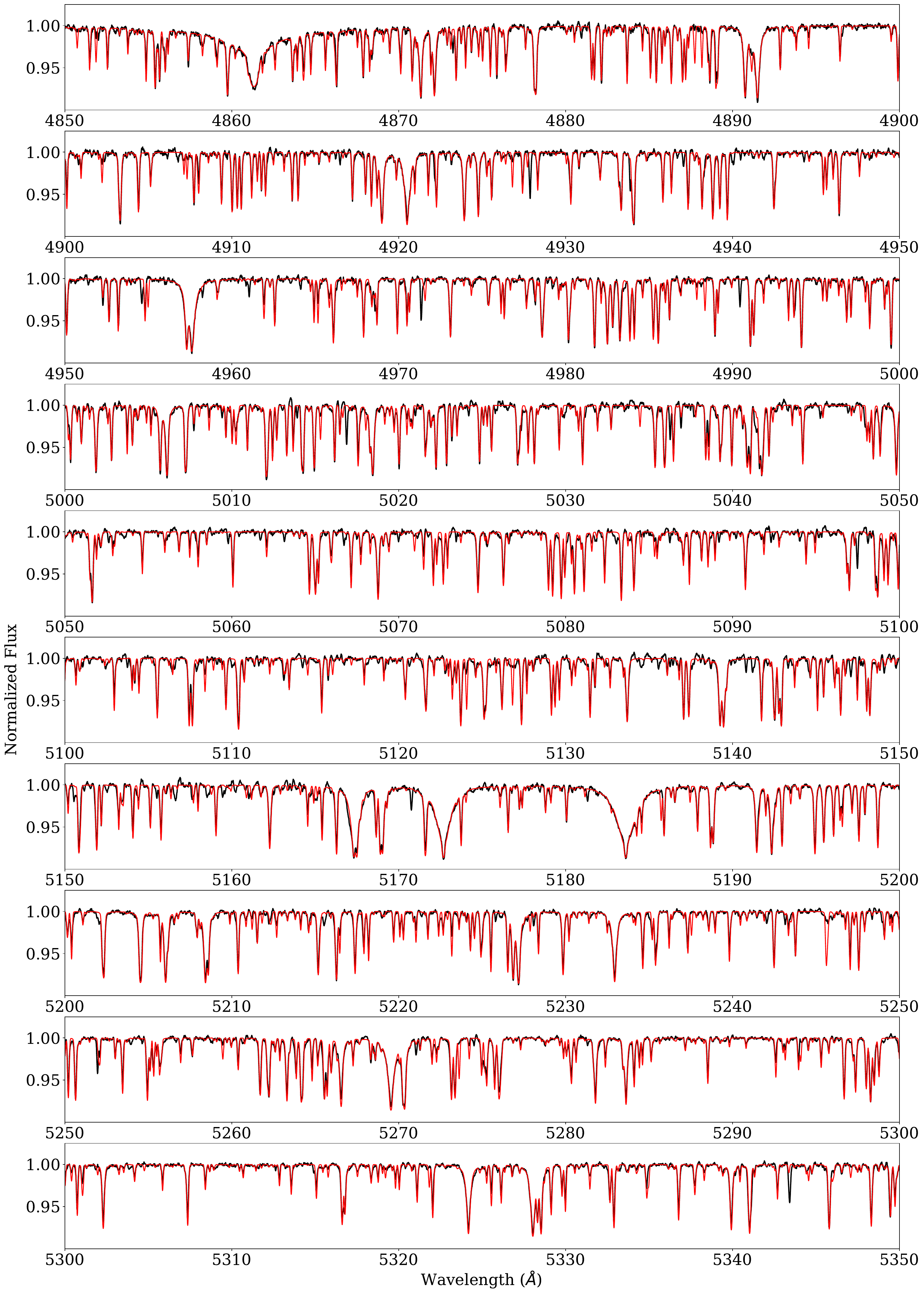}
    \caption{Mean spectrum of the secondary star in the CD-III range of PEPSI (4800-5400\,\AA). The black line is the observation, the red line is a model fit based on ParSES and MARCS.}
    \label{hr7275b_spec}
\end{figure*}

\begin{figure*}[h!]
    \centering
    \includegraphics[width= 12 cm]{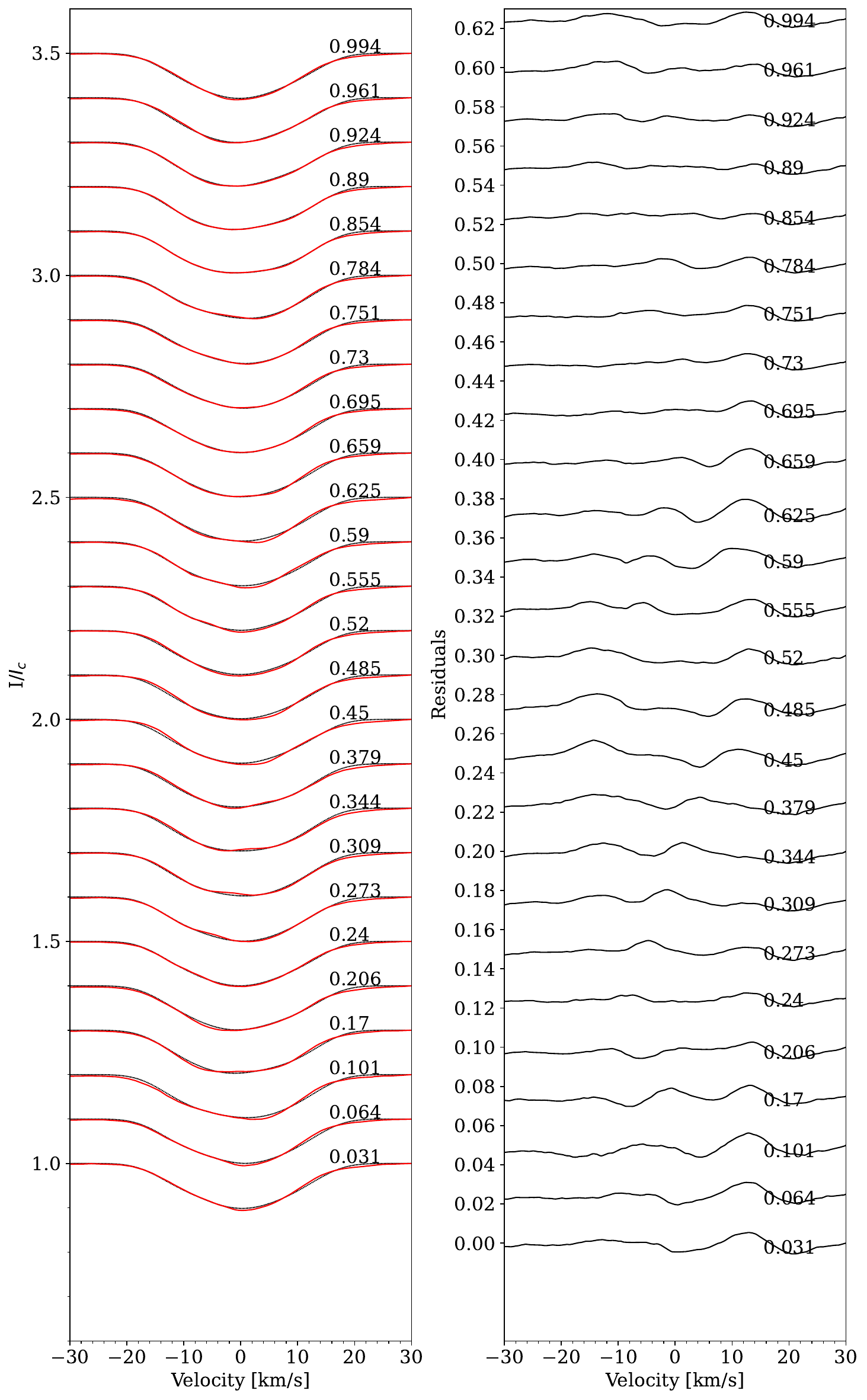}
    \caption{LSD profiles for the $i$MAP inversions. On the left panel the observed line profiles (red line) are compared with the fitted synthetic profiles (black dotted lines). The corresponding residuals are plotted in the right panel. Numbers on the right side of each panel indicate the rotational phase.}
    \label{SVD_prof}
\end{figure*}

\begin{figure*}[h!]
    \centering
    \includegraphics[width= \textwidth]{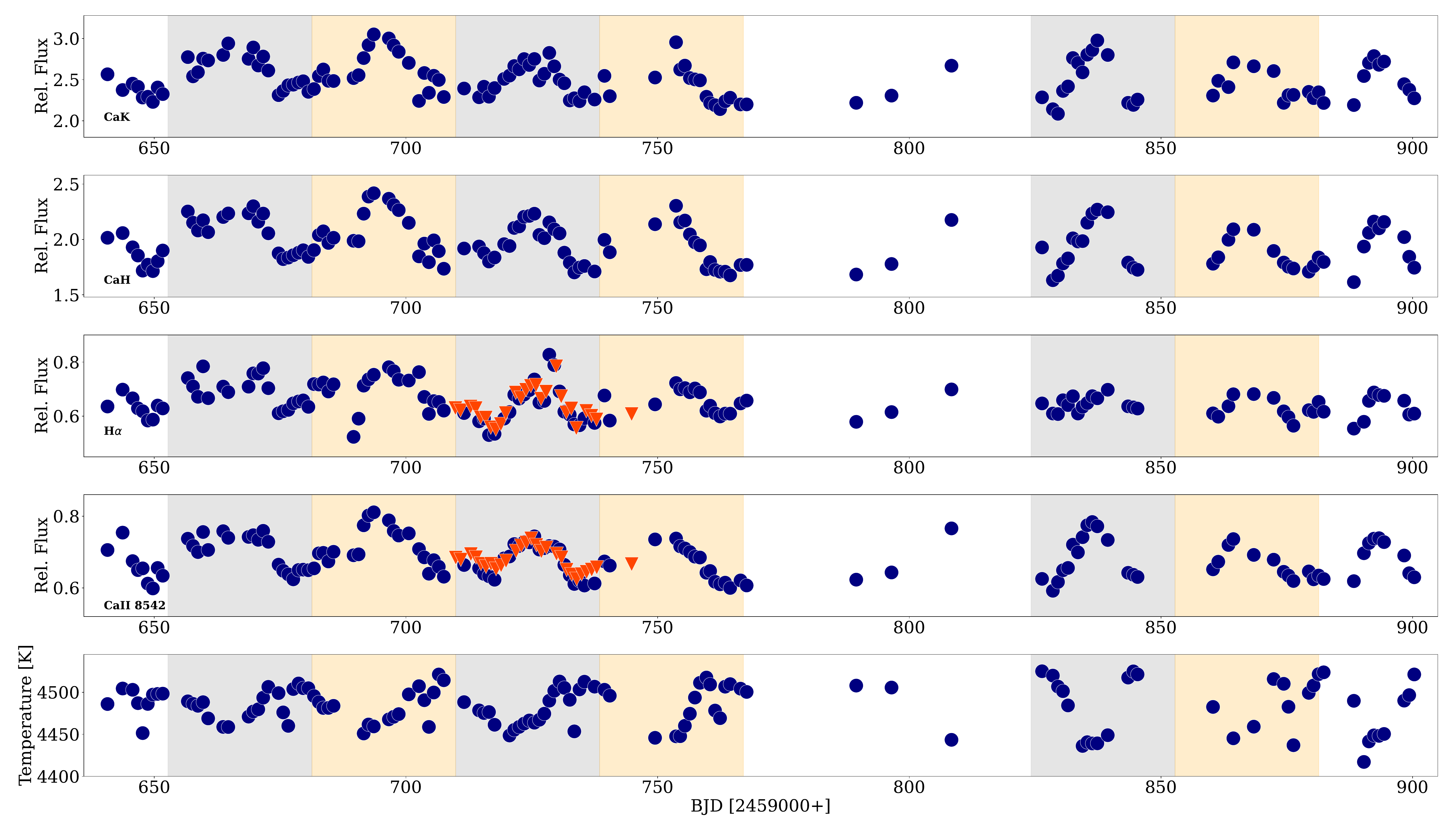}
    \caption{Chromospheric line emissions and effective temperature versus time. The dark blue circles indicate the STELLA observations while red triangles show the PEPSI data set. Consecutive complete rotations are shaded with gray and orange colors.}
    \label{Emissions}
\end{figure*}

\end{appendix}

\end{document}